\documentclass[12pt, draftclsnofoot, onecolumn]{IEEEtran}

\usepackage{fancyhdr}
\usepackage{amsmath}
\usepackage{amssymb}
\usepackage{graphicx}
\usepackage{booktabs}
\usepackage{amsfonts}
\usepackage{multirow}
\usepackage{algorithm}
\usepackage{algorithmic}
\usepackage{mathrsfs}
\usepackage{bm}
\usepackage{exscale}
\usepackage{relsize}
\usepackage{setspace}
\usepackage{color}
\usepackage{ulem}

\newtheorem{theorem}{\textbf{Theorem}}
\newtheorem{proposition}{\textbf{Proposition}}

\ifCLASSINFOpdf
\else
\fi

\begin{document}

\title{Spatial Modulation for More Spatial Multiplexing: RF-Chain-Limited Generalized Spatial Modulation Aided MmWave MIMO with Hybrid Precoding}

\author{
Longzhuang~He,~\IEEEmembership{Student~Member,~IEEE},
Jintao~Wang,~\IEEEmembership{Senior~Member,~IEEE},
and Jian~Song,~\IEEEmembership{Fellow,~IEEE}

\thanks{
Longzhuang He, Jintao Wang and Jian Song are with the Department of Electronic Engineering, Tsinghua University, Beijing, 100084, China (e-mail: helongzhuang@126.com; \{wangjintao, jsong\}@tsinghua.edu.cn).

This work was supported by the National Natural Science Foundation of China (Grant No. 61471221 and No. 61471219). Part of the material in this paper was submitted to the IEEE Global Communications Conference (Globecom), Singapore, Dec. 2017.
}
}

\maketitle
\begin{abstract}
The application of hybrid precoding in millimeter wave (mmWave) multiple-input multiple-output (MIMO) systems has been proved effective for reducing the number of radio frequency (RF) chains. However, the maximum number of independent data streams is conventionally restricted by the number of RF chains, which leads to limiting the spatial multiplexing gain. To further improve the achievable spectral efficiency (SE), in this paper we propose a novel generalized spatial modulation (GenSM) aided mmWave MIMO system to convey an extra data stream via the index of the active antennas group, while no extra RF chain is required. Moreover, we also propose a hybrid analog and digital precoding scheme for SE maximization. More specifically, a closed-form lower bound is firstly derived to quantify the achievable SE of the proposed system. By utilizing this lower bound as the cost function, a two-step algorithm is proposed to optimize the hybrid precoder. The proposed algorithm not only utilizes the concavity of the cost function over the digital power allocation vector, but also invokes the convex $\ell_\infty$ relaxation to handle the non-convex constraint imposed by analog precoding. Finally, the proposed scheme is shown via simulations to outperform state-of-the-art mmWave MIMO schemes in terms of achievable SE.
\end{abstract}


\IEEEpeerreviewmaketitle

\section{Introduction}
The concept of millimeter wave (mmWave) communication has been widely acknowledged to be an effective approach to substantially improve the system throughput for 5G telecommunication networks \cite{daniels2007emerging}-\cite{rangan2014mmWave}. More specifically, the available bandwidth of mmWave frequency ranging from $30$ to $300$ GHz is orders of magnitude wider than the available bandwidth in today's cellular networks operating in microwave bands, which is capable of enabling transmission rates of multi-gigabits per second (Gbps) and meet the $1,000$-fold capacity gain required by future 5G telecommunications \cite{huang2008milli}\cite{rappaport2013willwork}.

In order to compensate for the severe free-space pathloss of mmWave signals, mmWave communication is usually combined with multiple-input multiple-output (MIMO) systems and invokes precoding to overcome the pathloss and improve the signal-to-noise ratio (SNR) at the receiver end \cite{torkildson2011indoor}-\cite{zhou2015fast}. Conventional precoding schemes are usually operated entirely in the digital domain, i.e. a \textit{full-digital precoder}, in which each antenna is equipped with a dedicated radio frequency (RF) chain. Due to the high energy dissipation and cost of the RF chains \cite{hasan2011green}, the application of such full-digital precoding schemes can be quite disadvantageous.

In order to address this issue, it has been recently reported in \cite{ayach2014spatially}-\cite{gao2016energy} to employ the novel \textit{hybrid precoding} schemes for reducing the number of RF chains, in which a digital precoder is used to simultaneously adjust the transmitted symbols' phases and amplitudes, and an analog precoder is invoked for phase-shifting the RF-domain signals. More specifically, in \cite{ayach2014spatially} and \cite{chen2015iterative}, compressive sensing (CS) based approaches were exploited for the hybrid precoder designs, of which the performance was shown to be close to the full-digital waterfilling benchmark. Note that the schemes in \cite{ayach2014spatially} and \cite{chen2015iterative} employed a \textit{full-connected} architecture, i.e. each RF chain is simultaneously connected to all the antennas, which incurred high insertion loss and massive computational complexity in a massive MIMO context. To address this issue, in \cite{han2015large} and \cite{gao2016energy}, hybrid precoders with \textit{sub-connected} architectures were proposed, which provided a more favorable tradeoff between the hardware complexity and the achievable performance.

It is worth noting that, in the previous RF-chain-limited precoding schemes for mmWave MIMOs (including full-digital and hybrid precoding schemes), the maximum number of independent data streams available at the transmitter is restricted by the number of RF chains, which therefore limits the attainable spatial multiplexing (SMX) gain. To further explore the possibilities of increasing SMX gain in an RF-chain-limited mmWave system, mmWave MIMO has been recently combined with the concept of spatial modulation (SM) and generalized SM (GenSM) in \cite{liu2015SSK}-\cite{he2017spectral}. SM/GenSM is a novel extension of the conventional MIMO techniques, in which only a subset of antennas are randomly activated by the input information to transmit the classic amplitude-phase modulation (APM) symbols \cite{mesleh2008spatial}-\cite{he2015infinity}. The information in SM/GenSM systems is not only transmitted by the APM symbols (APM-domain information), but is also conveyed by the indices of the active antennas (space-domain information). As the space-domain information does not require an extra RF chain, it is thus possible to employ SM/GenSM for improving the achievable spectral efficiency (SE) of RF-chain-limited mmWave MIMOs.

More specifically, in \cite{liu2015SSK} and \cite{liu2016LOS}, the applications of space shift keying (SSK) \cite{jeganathan2009SSK} and GenSM for indoor line-of-sight (LoS) channels were investigated, where the authors proposed to elaborately design the spacing of the antennas for performance optimization. In \cite{ishikawa2016GSM}, the application of analog beamforming (ABF) in GenSM-aided mmWave MIMO systems was explored. It was shown by \cite{ishikawa2016GSM} that, aided with ABF, the constrained capacity of the proposed system had the potential to approach the unprecoded MIMO capacity in low SNR regions, while maintaining a reduced-RF-chain structure. However, the preceding research on GenSM-aided mmWave MIMOs \cite{liu2015SSK}-\cite{ishikawa2016GSM} all failed to fully exploit the transmitter's knowledge of the channel state information (CSI), hence their achievable rates were far worse than the optimal MIMO capacity achieved by waterfilling precoding \cite{perez2010MIMO}. Although \cite{he2017spectral} considered the issue of analog precoding for GenSM-aided mmWave MIMO, the performance was still far from optimal due to the lack of digital precoding.

In fact, to the best of the authors' knowledge, the design of hybrid precoding in a GenSM-aided mmWave MIMO scenario has not been explored yet. Therefore it is of paramount importance to develop an efficient hybrid precoding scheme for GenSM-aided mmWave MIMOs in terms of SE maximization.

In this context, the major contributions of our paper can be summarized as follows.

\begin{enumerate}
  \item We extend the sub-connected hybrid precoding structures originally proposed in \cite{han2015large}\cite{gao2016energy}, and propose our novel GenSM-aided mmWave MIMO scheme. Different from the conventional mmWave schemes, an extra data stream can be modulated in our system without requiring any extra RF chains, which leads to increasing the degrees of spatial freedom. More importantly, our proposed system is a more generalized sub-connected mmWave MIMO structure, and the conventional sub-connected structures in \cite{han2015large} and \cite{gao2016energy} are conceived as special cases when the space-domain information transmission is removed from our structure.

  \item Due to the prohibitive complexity required for evaluating the achievable SE of the proposed system, in this paper we propose a closed-form SE lower bound, which significantly reduces the computational complexity for SE analysis. The proposed SE bound is also shown to provide an accurate approximation to the true SE, when a constant shift is applied.

  \item By utilizing the proposed bound as a low-complexity cost function, we propose a two-step algorithm to design the digital and analog precoders. More importantly, as the proposed bound is proved to be a concave function of the digital precoder's power allocation vector, the digital precoder is therefore designed within the framework of convex optimization. The optimization of the analog precoder's coefficients is originally a problem with a non-convex constraint, which is relaxed to a convex $\ell_\infty$ constraint and solved via a gradient ascent method.

  \item As the conventional hybrid precoding schemes with sub-connected structures can be treated as special cases of the proposed framework (when the space-domain information transmission is removed), potential SE gain is thus achievable by the proposed, more generalized structure. In fact, by optimizing the system configuration parameters, substantial SE gain are observed via numerical simulations.

\end{enumerate}

Note that part of the material of this paper has been submitted for peer review in \cite{he2017generalized}. The major difference between this manuscript and \cite{he2017generalized} are:
\begin{enumerate}
  \item The theoretical derivations and mathematical proofs of the theorems and propositions are all presented in this paper, which were omitted for brevity in \cite{he2017generalized}.
  \item The proposed algorithm for the hybrid precoder design is introduced with more technical details in this paper, as opposed to \cite{he2017generalized}.

  \item In this paper, we also discuss the convergence of the proposed algorithm as well as its dependence on the initial points, which were again absent in \cite{he2017generalized}. We found that the proposed two-step algorithm is very robust to the variation of initial points.

  \item The complexity analysis is conducted in this paper, while the parameter optimization is also presented with more simulation results in Section \uppercase\expandafter{\romannumeral4}-F of this paper.
\end{enumerate}

Therefore this paper should be treated as a substantial extension of the content in \cite{he2017generalized}.

The organization of this paper is introduced as follows. Section \uppercase\expandafter{\romannumeral2} introduces the system model of our proposed GenSM-aided mmWave MIMO. Theoretical SE analysis is provided in Section \uppercase\expandafter{\romannumeral3}. Section \uppercase\expandafter{\romannumeral4} introduces our proposed two-step optimization algorithm. The simulation and comparison results are provided in Section \uppercase\expandafter{\romannumeral5}, while Section \uppercase\expandafter{\romannumeral6} concludes this paper.


\textit{Notations}: The lowercase and uppercase boldface letters denote column vectors and matrices respectively. The operators $(\cdot)^T$ and $(\cdot)^H$ denote the transposition and conjugate transposition, respectively. $\mathcal{CN}(\bm\mu, \bm\Sigma)$ denotes a circularly symmetric complex-valued multi-variate Gaussian distribution with $\bm\mu$ and $\bm\Sigma$ being its mean and covariance, respectively, while $\mathcal{CN}(\mathbf{x}; \bm\mu, \bm\Sigma)$ denotes the probability density function (PDF) of a random vector $\mathbf{x} \sim \mathcal{CN}(\bm\mu, \bm\Sigma)$. $\mathbf{M}_{(i, j)}$ is used to denote the $(i; j)$ component of a matrix $\mathbf{M}$. $\|\mathbf{M}\|_F$ represents the Frobenius norm of $\mathbf{M}$ and $\vert \mathbf{M} \vert$ is the determinant. $\mathbf{I}_N$ denotes an $N$-dimensional identity matrix, and $\mathbf{e}_n \in \mathbb{R}_{N \times 1}$ represents the $n$-th column of $\mathbf{I}_N$. The $\ell_\infty$ norm of a vector $\mathbf{a} \in \mathbb{C}_{N \times 1}$ is defined as $\|\mathbf{a}\|_\infty = \max_{n=1, \ldots, N} |a_n|$.

\section{System Model}
\begin{figure}
\center{\includegraphics[width=0.75\linewidth]{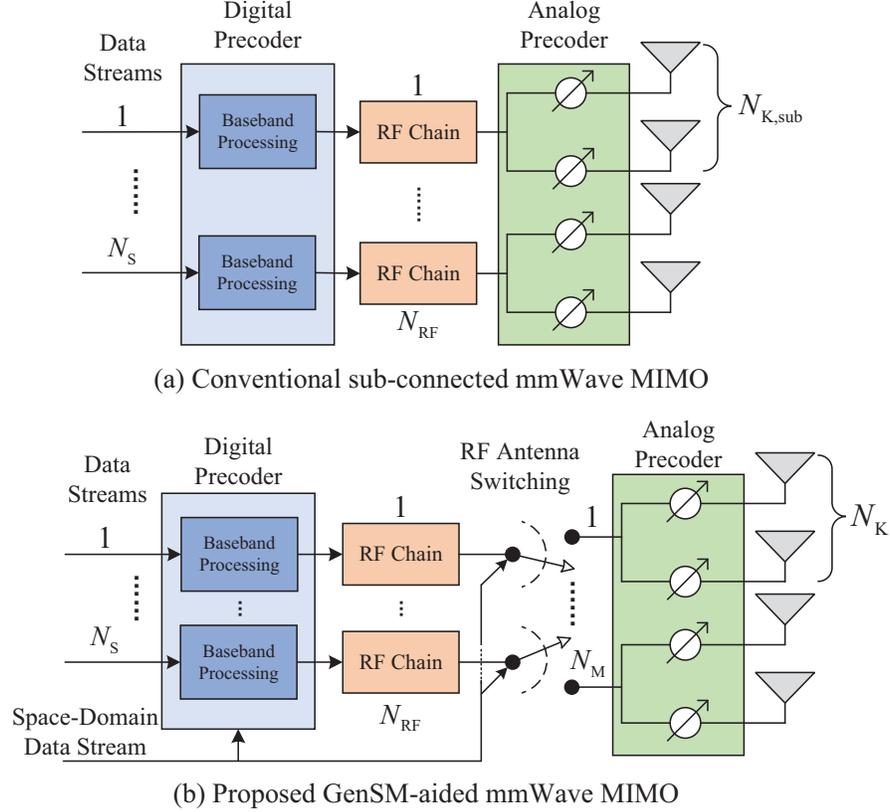}}
\caption{Block diagrams of (a) conventional sub-connected mmWave MIMO scheme and (b) proposed GenSM-aided mmWave MIMO scheme.}
\label{Fig_SystemModel}
\end{figure}

The proposed GenSM-aided mmWave MIMO can be considered as a combination of the sub-connected mmWave structure \cite{han2015large}\cite{gao2016energy} and the information-guided antenna-switching principle of GenSM. In order to provide an intuitive demonstration, in Fig.\ref{Fig_SystemModel} we provide the block diagrams of the conventional sub-connected mmWave MIMO scheme and our proposed scheme. For the conventional sub-connected mmWave MIMO scheme in Fig.\ref{Fig_SystemModel} (a), it can be seen that $N_\text{S}$ independent data streams are firstly processed by a diagonal digital precoder, which essentially plays the role of \textit{power allocation}. After the digital precoder, $N_\text{RF}$ RF symbols are generated with $N_\text{RF}$ denoting the number of RF chains. The output of each RF chain is then assigned to $N_{\text{K,sub}}$ phase shifters (PSs) for analog precoding. In this paper we denote the number of transmit antennas (TAs) and receive antennas (RAs) as $N_\text{T}$ and $N_\text{R}$, respectively. Therefore we have $N_\text{T} = N_\text{RF} N_\text{K,sub}$ for the conventional sub-connected scheme in Fig.\ref{Fig_SystemModel} (a). According to Fig.\ref{Fig_SystemModel} (a), it is required by the conventional sub-connected structure that $N_\text{S} = N_\text{RF}$, hence the potentially attainable SMX gain is restricted by the number of RF chains.

In order to address this problem, in our proposed GenSM-aided mmWave MIMO system depicted in Fig.\ref{Fig_SystemModel} (b), an extra data stream, i.e. the space-domain data stream, is also modulated in the transmitted signal. More specifically, the $N_\text{S}$ data streams are also processed with a diagonal digital precoder. Different from Fig.\ref{Fig_SystemModel} (a), the power allocation vector of the digital precoder is simultaneously determined by the space-domain information input and the instantaneous CSI. Thanks to the high-speed and low-latency advantages of the baseband digital processing, this space-information-guided digital precoding can be performed for each symbol's transmission. Moreover, the $N_\text{T}$ TAs are divided into $N_\text{M}$ antenna groups (AGs), each of which consists of $N_\text{K}$ TAs, hence we have $N_\text{T} = N_\text{M} N_\text{K}$. In our proposed system, it is required that $N_\text{M} \ge N_\text{RF}$, and the space-domain information can therefore randomly assign the outputs of the $N_\text{RF}$ RF chains to $N_\text{RF}$ out of the $N_\text{M}$ AGs, while the remaining $(N_\text{M} - N_\text{RF})$ AGs are kept silent during this symbol's transmission. Similar to Fig.\ref{Fig_SystemModel} (a), $N_\text{K}$ PSs are also invoked in each AG to perform analog precoding. Finally, it is worth noting that the conventional sub-connected mmWave MIMO is actually a special case of our proposed scheme when $N_\text{M} = N_\text{RF}$, hence the proposed system is a more generalized version of the sub-connected mmWave MIMO.

We let $\mathbf{x} \in \mathbb{C}_{N_\text{S} \times 1}$ represent the transmitted symbol vector, which is assumed to distribute as $\mathbf{x} \sim \mathcal{CN}(\mathbf{0}, \frac{1}{N_\text{S}} \mathbf{I}_{N_\text{S}})$, according to \cite{ayach2014spatially}\cite{gao2016energy}. As the space-domain information plays the role of selecting an active AGs' combination (AGC), therefore the total number of legitimate AGCs, i.e. $M$, can be given as \cite{wang2012generalised}:
\begin{equation}
  \displaystyle M = 2^{\left\lfloor \log_2 \binom{N_\text{M}}{N_\text{RF}} \right\rfloor},
  \label{M_Definition}
\end{equation}
where $\lfloor \cdot \rfloor$ represents the floor operation, and $\binom{\cdot}{\cdot}$ represents the binomial coefficient. Moreover, we use $\mathbf{u}_m \triangleq [u_{m1}, u_{m2}, \ldots, u_{m N_\text{RF}} ]^T $ to denote the indices of the AGs activated by the $m$-th AGC ($m = 1, 2, \ldots, M$), which are subject to the following ordering constraint:
\begin{equation}
  1 \le u_{m1} < u_{m2} < \ldots < u_{m N_\text{RF}} \le N_\text{M}.
\end{equation}

Hence the $m$-th \textit{AG-selection matrix} $\mathbf{C}_m \in \mathbb{R}_{N_\text{T} \times N_\text{RF}}$ can be defined as follows ($1 \le m \le M$):
\begin{equation}
  \mathbf{C}_m \triangleq \left[\mathbf{e}_{u_{m1}}, \mathbf{e}_{u_{m2}}, \ldots, \mathbf{e}_{u_{mN_\text{RF}}}\right] \otimes \mathbf{1}_{N_\text{K}},
\end{equation}
where $\mathbf{1}_{N_\text{K}} \in \mathbb{R}_{N_\text{K} \times 1}$ denotes an $N_\text{K}$-dimensional all-one vector, $\mathbf{e}_m$ represents the $m$-th column of $\mathbf{I}_{N_\text{M}}$ with $1 \le m \le N_\text{M}$, and $\otimes$ represents the Kronecker product. Note that, for each symbol's transmission, the space-domain information randomly selects one of the $M$ AGCs according to a \textit{uniform probability distribution}.

Moreover, the digital precoder is specified to be a collection of $M$ diagonal real-valued matrices, i.e. $\mathcal{D} \triangleq \left\{\mathbf{D}_1, \ldots, \mathbf{D}_M\right\}$, where $\mathbf{D}_m = \text{diag}\left(d_{m1}, d_{m2}, \ldots, d_{mN_\text{S}}\right) \in \mathbb{R}_{N_\text{S} \times N_\text{S}}$ is the applied precoder when the $m$-th AGC is selected by the space-domain information.

As the analog precoder plays the role of phase-shifting, the corresponding precoder matrix can thus be denoted as $\mathbf{A} \in \mathbb{C}_{N_\text{T} \times N_\text{T}}$ and given by:
\begin{equation}
  \mathbf{A} \triangleq \displaystyle \text{diag}\left( \frac{1}{\sqrt{N_\text{K}}} e^{j\theta_1}, \frac{1}{\sqrt{N_\text{K}}} e^{j\theta_2}, \ldots, \frac{1}{\sqrt{N_\text{K}}} e^{j\theta_{N_\text{T}}}  \right),
  \label{AForm}
\end{equation}
where $\theta_{n} \in [0, 2\pi)$ denotes the rotation phase of the $n$-th TA.

Finally, the received signal vector $\mathbf{y} \in \mathbb{C}_{N_\text{R} \times 1}$ at the receiver end, when the $m$-th AGC is selected, can thus be formulated as:
\begin{equation}
  \mathbf{y} = \displaystyle \sqrt{\rho} \mathbf{H} \mathbf{A} \mathbf{C}_m \mathbf{D}_m \mathbf{x} + \mathbf{n},
  \label{ReceivedSignal}
\end{equation}
where a narrowband MIMO channel matrix $\mathbf{H} \in \mathbb{C}_{N_\text{R} \times N_\text{T}}$ is considered as in \cite{gao2016energy}\cite{ishikawa2016GSM}. Similar to \cite{ayach2014spatially}, the power of $\mathbf{H}$ has been normalized so that $E\left\{\|\mathbf{H}\|_F^2\right\} = N_\text{R} N_\text{T}$. The average transmit power is given by $\rho > 0$, while $\mathbf{n} \sim \mathcal{CN}(\mathbf{0}, \sigma_\text{N}^2 \mathbf{I}_{N_\text{R}})$ represents the additive white Gaussian noise (AWGN) at the receiver side. In order to maintain an SNR value $\rho / \sigma_\text{N}^2$ at the receiver, the digital precoder $\mathcal{D}$ must satisfy the following power constraint:
\begin{equation}
    \displaystyle \sum_{m=1}^M \text{Tr}\left(\mathbf{D}_m \mathbf{D}_m^H \right) \le M N_\text{S}.
    \label{FConstraint}
\end{equation}

Similar to \cite{ayach2014spatially} and \cite{gao2016energy}, in this paper we adopt the classic clustered Saleh-Valenzuela mmWave channel model, which is formulated as:
\begin{equation}
    \mathbf{H} = \displaystyle \gamma \sum_{p=1}^{N_\text{cl}} \sum_{q=1}^{N_\text{ray}} \alpha_{pq} \Lambda_\text{t}(\phi_{pq}^\text{t}, \theta_{pq}^\text{t}) \Lambda_\text{r}(\phi_{pq}^\text{r}, \theta_{pq}^\text{r}) \mathbf{b}_\text{t}(\phi_{pq}^\text{t}, \theta_{pq}^\text{t}) \mathbf{b}_\text{r}(\phi_{pq}^\text{r}, \theta_{pq}^\text{r}),
    \label{SVChannel}
\end{equation}
where $\gamma > 0$ is the normalizing factor ensuring $E\{\|\mathbf{H}\|_F^2\} = N_\text{R} N_\text{T}$, $N_\text{cl}$ is the number of scattering clusters, and $N_\text{ray}$ denotes the number of effective propagation paths within each cluster. The complex-valued channel gain is given by $\alpha_{pq} \sim \mathcal{CN}(0, \sigma_{\alpha,p}^2)$. Moreover, the azimuth (elevation) angles of departure and arrival (AoDs and AoAs) at the transmitter and the receiver are given by $\phi_{pq}^\text{t} (\theta_{pq}^\text{t})$ and $\phi_{pq}^\text{r} (\theta_{pq}^\text{r})$, respectively. The transmit and receive antenna gains are denoted by $\Lambda_\text{t}(\phi_{pq}^\text{t}, \theta_{pq}^\text{t})$ and $\Lambda_\text{r}(\phi_{pq}^\text{r}, \theta_{pq}^\text{r})$, respectively, while $\mathbf{b}_\text{t}(\phi_{pq}^\text{t}, \theta_{pq}^\text{t})$ and $\mathbf{b}_\text{r}(\phi_{pq}^\text{r}, \theta_{pq}^\text{r})$ denote the normalized transmit and receive antenna array responses given by \cite{ayach2014spatially}:
\begin{equation}
  \mathbf{b}_\tau(\phi_{pq}^\tau, \theta_{pq}^\tau) = \displaystyle \frac{1}{\sqrt{U}}\left[ 1, e^{j\frac{2\pi}{\lambda} d \sin\left(\phi_{pq}^\tau\right)}, \ldots, e^{j(U-1)\frac{2\pi}{\lambda} d \sin\left(\phi_{pq}^\tau\right)} \right]^T,
  \label{ArrayResponse}
\end{equation}
with $\tau \in \{\text{t}, \text{r}\}$. The number and spacing of the antenna elements are given by $U$ and $d$, while $\lambda$ represents the signal's wavelength. As we assume that the transmit and receive antennas form two horizontal uniform linear arrays (ULAs), (\ref{ArrayResponse}) is therefore irrelevant to the elevation angles $\theta_{pq}^\tau$. Similar to \cite{ayach2014spatially}, the angles $\phi_{pq}^\tau$ ($\theta_{pq}^\tau$) are assumed to be Laplacian random variables with a uniformly-random mean cluster angle $\bar{\phi}_p^\tau$ ($\bar{\theta}_p^\tau $) and angle spread $\sigma_\phi^\tau$ ($\sigma_\theta^\tau$). Lastly, we assume that the antenna element gains are given as \cite{ayach2014spatially}:
\begin{equation}
  \Lambda_\tau(\phi_{pq}^\tau, \theta_{pq}^\tau) =
  \begin{cases}
    1, & \phi_{pq}^\tau \in [\phi_\text{min}^\tau, \phi_\text{max}^\tau], \\
    0, & \text{otherwise},
  \end{cases}
\end{equation}
where $[\phi_\text{min}^\tau, \phi_\text{max}^\tau]$ are the azimuth sector angles at the transmitter ($\tau = \text{t}$) and receiver ($\tau = \text{r}$).

It is worth noting that, the model in (\ref{SVChannel}) is certainly not the only channel model suitable for the analysis and algorithms of this paper. As this paper is more concerned about the specific channel realization $\mathbf{H}$, hence the LoS channel model \cite{liu2016LOS} or the 3-D mmWave channel model \cite{samimi20163d} would also be applicable.

Finally, in order that the $N_\text{S}$ independent data streams can be successfully transmitted, in this paper we require that
\begin{equation}
  N_\text{S} \le \text{rank}(\mathbf{H}).
  \label{rankRequirement}
\end{equation}

\section{Theoretical Spectral Efficiency Analysis}
\subsection{Mutual Information Analysis}
According to (\ref{ReceivedSignal}), the achievable SE of the proposed system can be characterized via the mutual information (MI) between $\mathbf{y}$, $\mathbf{x}$ and $m$, i.e.
\begin{equation}
  R(\mathbf{H}, \mathcal{D}, \mathbf{A}) = I(\mathbf{y}; \mathbf{x}, m),
  \label{MI0}
\end{equation}
of which the left-hand side indicates that the MI term is a function of the instantaneous channel realization $\mathbf{H}$, the digital precoder $\mathcal{D}$ and the analog precoder $\mathbf{A}$.

Due to the discrete-random channel input $m \in \{1, 2, \ldots, M\}$, the MI term in (\ref{MI0}) cannot be expressed in a \textit{closed form}, and it can only be obtained via numerical integrations, which requires prohibitive complexity \cite{he2017on}\cite{za2014mianalysis}. Therefore we propose Theorem \ref{theorem1} to provide a closed-form expression $R_\text{LB}(\mathbf{H}, \mathcal{D}, \mathbf{A})$ for lower-bounding $R(\mathbf{H}, \mathcal{D}, \mathbf{A})$:
\begin{theorem}
  A closed-form lower bound for the achievable SE of the proposed system is given as follows:
  \begin{equation}
    R_\text{LB}(\mathbf{H}, \mathcal{D}, \mathbf{A}) = \displaystyle \log_2 \frac{M}{(e \sigma_\text{N}^2)^{N_\text{R}}} - \frac{1}{M} \sum_{n=1}^M \log_2 \sum_{t=1}^M\left|\bm\Sigma_n + \bm\Sigma_t\right|^{-1},
    \label{MILB}
  \end{equation}
  where $\bm\Sigma_n$ is given as follows:
  \begin{equation}
    \bm\Sigma_n \triangleq \sigma_\text{N}^2\mathbf{I}_{N_\text{R}} + \frac{\rho}{N_\text{S}} \mathbf{HA}\mathbf{C}_n \mathbf{D}_n \mathbf{D}_n^H \mathbf{C}_n^H \mathbf{A}^H \mathbf{H}^H.
    \label{SigmanDef}
  \end{equation}
  \label{theorem1}
\end{theorem}

\begin{IEEEproof}
  The proof is provided in the Appendix \ref{AppA}.
\end{IEEEproof}

The closed-form lower bound $R_\text{LB}$ proposed by Theorem \ref{theorem1} has facilitated a computationally efficient approach to quantify the achievable SE performance. We now move on to demonstrate the tightness of the proposed bound $R_\text{LB}$.

\subsection{Bound Tightness}
We now use several examples to demonstrate the tightness of the proposed closed-form bound $R_\text{LB}(\mathbf{H}, \mathcal{D}, \mathbf{A})$. Before presenting the numerical results, we firstly propose the following proposition to discuss the issue of bound tightness.
\begin{proposition}
  A constant gap of $N_\text{R}(1 - \log_2 e)$ exists between $R_\text{LB}(\mathbf{H}, \mathcal{D}, \mathbf{A})$ and the true SE expression $R(\mathbf{H}, \mathcal{D}, \mathbf{A})$, when an asymptotically high or low SNR is imposed.
  \label{proposition1}
\end{proposition}

\begin{IEEEproof}
  The proof is provided in the Appendix \ref{AppB}.
\end{IEEEproof}

\begin{table*}
\small
\caption{Simulation Parameters}
\newcommand{\tabincell}[2]{\begin{tabular}{@{}#1@{}}#2\end{tabular}}
\centering
\renewcommand\arraystretch{1.2}
\begin{tabular}{l|l|l}
\hline\hline
Symbols                 & Specifications & Typical Values \\\hline\hline
$N_\text{T}$            & Number of TAs                                             & $8$ \\\hline
$N_\text{R}$            & Number of RAs                                             & $8$ \\\hline
$N_\text{K}$            & Number of TAs in each antenna group                       & $2$ \\\hline
$N_\text{M}$            & Number of antenna groups                                  & $4$ \\\hline
$N_\text{RF}$           & Number of RF chains                                       & $2$ \\\hline
$N_\text{S}$            & Number of APM-domain data streams                         & $N_\text{RF}$ \\\hline
$\lambda$               & Carrier's wavelength                                      & $5$ mm \\\hline
$N_\text{cl}$           & Number of scattering clusters                             & $8$   \\\hline
$N_\text{ray}$          & Number of propagation paths                               & $10$  \\\hline
$\sigma_{\alpha, p}^2$  & Average power of the $p$-th cluster                       & $1$   \\\hline
$\sigma_{\phi}^\tau (\sigma_{\theta}^\tau)$ & Azimuth (elevation) angular spreads, $\tau \in \{\text{t}, \text{r}\}$    & $7.5^\circ$ \\\hline
$[\phi_\text{min}^\text{t}, \phi_\text{max}^\text{t}]$  & Azimuth sector angles at the transmitter  & $[-30^\circ, 30^\circ]$ \\\hline
$[\phi_\text{min}^\text{r}, \phi_\text{max}^\text{r}]$  & Azimuth sector angles at the receiver  & $[-180^\circ, 180^\circ]$ \\\hline
\hline
\end{tabular}
\label{TABLESimu}
\end{table*}

Therefore we can apply this constant shift in $R_\text{LB}(\mathbf{H}, \mathcal{D}, \mathbf{A})$ to obtain a more accurate, asymptotically unbiased approximation to $R(\mathbf{H}, \mathcal{D}, \mathbf{A})$. Before presenting our results on the bound tightness, we summarize the typical values of simulation parameters in Table \ref{TABLESimu} and stress that all the simulations in this paper are configured according to Table \ref{TABLESimu}, unless mentioned otherwise. As the specific design of $\mathcal{D}$ and $\mathbf{A}$ has not yet been discussed, we therefore apply the trivial precoding scheme, i.e.
\begin{equation}
  \mathcal{D} = \left\{\mathbf{I}_{N_\text{S}}, \ldots, \mathbf{I}_{N_\text{S}}\right\}, \,\, \mathbf{A} = \frac{1}{\sqrt{N_\text{K}}} \mathbf{I}_{N_\text{T}}.
\end{equation}

\begin{figure}
\center{\includegraphics[width=0.75\linewidth]{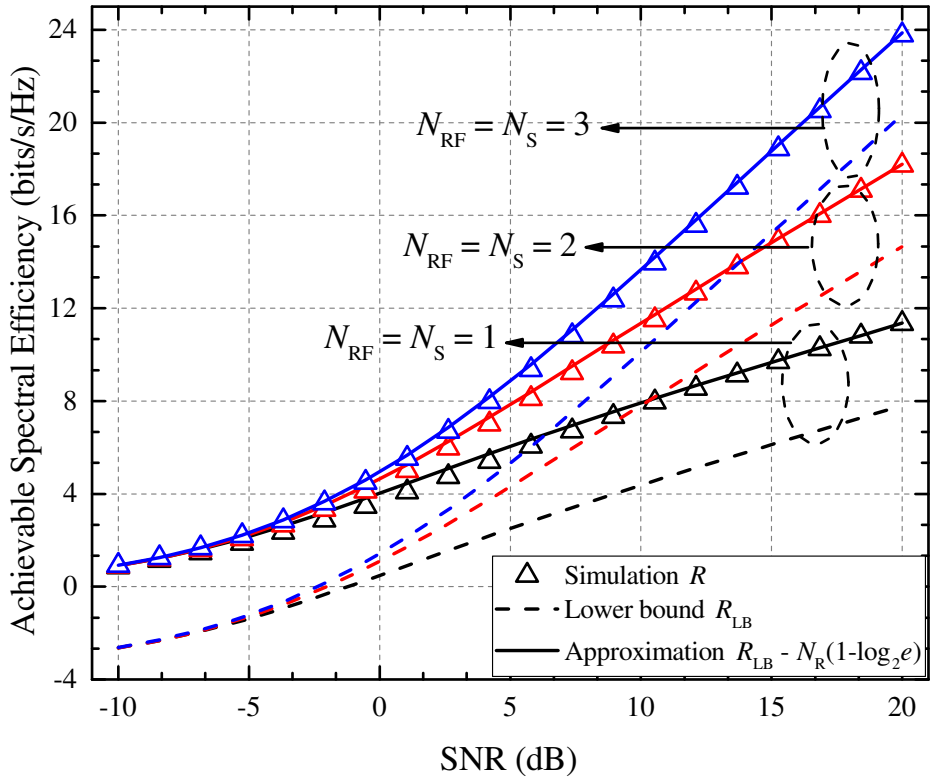}}
\caption{Average achievable SE and SE bounds with $N_\text{S} = N_\text{RF} = 1$, $2$ and $3$. The simulation value $R$ and closed-form lower bound $R_\text{LB}$ are given according to (\ref{MI0}) and (\ref{MILB}), respectively. The approximation is given by $R_\text{LB} - N_\text{R}(1 - \log_2 e)$. Other parameters are specified according to Table \ref{TABLESimu}.}
\label{Fig_BoundTightness}
\end{figure}

In Fig.\ref{Fig_BoundTightness}, the true SE expression $R(\mathbf{H}, \mathcal{D}, \mathbf{A})$ as well as the SE lower bound $R_\text{LB}(\mathbf{H}, \mathcal{D}, \mathbf{A})$ (with and without constant shift) averaged over $2,000$ random channel realizations are depicted as a function of SNR $\rho / \sigma_\text{N}^2$. As it can be seen from the figure, although the proposed lower bound $R_\text{LB}$ exhibits an SE gap with respect to the true SE $R$, it actually provides a favorable approximation accuracy when the constant shift $N_\text{R}(1 - \log_2 e)$ is compensated. Since adding the constant shift imposes no impact on the precoder design in terms of SE maximization, we would therefore utilize $R_\text{LB}$ as a low-complexity cost function to design the hybrid precoders in the following sections.

\section{Proposed Two-Step Algorithm For Precoder Design}
In this section, we propose to solve the following optimization problem (P1) to design the hybrid precoders:
\begin{equation}
\arraycolsep=1.0pt\def\arraystretch{1.3}
  \begin{array}{lcl}
  \text{(P1):    }  & \displaystyle \max_{\mathcal{D}, \mathbf{A}}  & \displaystyle R_\text{LB}(\mathbf{H}, \mathcal{D}, \mathbf{A}) \\
                    & \displaystyle \text{s.t.}                     & \mathcal{D} = \left\{\mathbf{D}_1, \mathbf{D}_2, \ldots, \mathbf{D}_M\right\}, \\
                    &                                               & \displaystyle \sum_{m=1}^M \text{Tr}(\mathbf{D}_m \mathbf{D}_m^H) \le M N_\text{S} \,\, \text{and} \,\,  \mathbf{A} \in \mathscr{A},
  \end{array}
  \label{P1}
\end{equation}
where $\mathscr{A}$ represents the feasible set of $\mathbf{A}$ satisfying the definition in (\ref{AForm}).

In order to solve (P1) with a reduced level of complexity, similar to \cite{zeng2012linear}, we propose to decompose (P1) and \textit{iteratively} solve the following two sub-problems, i.e. (P2) and (P3):
\begin{equation}
\arraycolsep=1.0pt\def\arraystretch{1.3}
  \begin{array}{lcl}
  \text{(P2):    }  & \displaystyle \max_{\mathcal{D}}              & \displaystyle R_\text{LB}(\mathbf{H}, \mathcal{D}, \mathbf{A}) \\
                    & \displaystyle \text{s.t.}                     & \mathcal{D} = \left\{\mathbf{D}_1, \mathbf{D}_2, \ldots, \mathbf{D}_M\right\}, \\
                    &                                               & \displaystyle \sum_{m=1}^M \text{Tr}(\mathbf{D}_m \mathbf{D}_m^H) \le M N_\text{S}, \\
  \text{(P3):    }  & \displaystyle \max_{\mathbf{A}}               & \displaystyle R_\text{LB}(\mathbf{H}, \mathcal{D}, \mathbf{A}) \,\, \text{s.t.} \,\, \mathbf{A} \in \mathscr{A},
  \end{array}
\end{equation}
in which (P2) optimizes the digital precoder $\mathcal{D}$ based on a given $\mathbf{A}$, while (P3) optimizes the analog precoder $\mathbf{A}$ upon assuming an invariable $\mathcal{D}$. In the following subsections we will introduce our solutions to (P2) and (P3), as well as the final proposed two-step algorithm.

\subsection{Digital Precoder Design for SE Maximization}
\begin{figure*}
\small
    \begin{equation}
    \arraycolsep=1.0pt\def\arraystretch{3.0}
    \begin{array}{rcl}
      \displaystyle\nabla_{\bm\lambda_m} R_\text{LB}(\bm\lambda) &=& \displaystyle \frac{\rho \log_2 e}{MN_\text{S}} \sum_{n=1}^M \frac{\left|\bm\Sigma_n + \bm\Sigma_m\right|^{-1} \text{diag}\left[ \mathbf{C}_m^H\mathbf{A}^H \mathbf{H}^H  \left(\bm\Sigma_n + \bm\Sigma_m\right)^{-1} \mathbf{HAC}_m \right]}{\sum_{t=1}^M \left| \bm\Sigma_n + \bm\Sigma_t \right|^{-1}} + \text{...} \\
      && \displaystyle \frac{\rho \log_2 e}{MN_\text{S}} \frac{\sum_{t=1}^M \left| \bm\Sigma_m + \bm\Sigma_t \right|^{-1} \text{diag}\left[\mathbf{C}_m^H \mathbf{A}^H \mathbf{H}^H \left( \bm\Sigma_m + \bm\Sigma_t \right)^{-1} \mathbf{HAC}_m\right] }{\sum_{t=1}^M \left| \bm\Sigma_m + \bm\Sigma_t \right|^{-1}}.
    \end{array}
    \label{RLB_Lambda_m}
    \end{equation}
\hrulefill
\end{figure*}

The solution to (P2) essentially relies on the concavity of $R_\text{LB}$ over $\mathbf{D}_m$. Unfortunately such concavity does not hold, which can be readily verified by a counterexample (e.g. $N_\text{R} = N_\text{T} = N_\text{K} = N_\text{M} = N_\text{RF} = M = 1$).

Since $\mathbf{D}_m = \text{diag}(d_{m1}, \ldots, d_{mN_\text{S}})$ is a diagonal matrix, another option is to verify the concavity of $R_\text{LB}$ with respect to $\bm\lambda_m \triangleq [d_{m1}^2, d_{m2}^2, \ldots, d_{m N_\text{S}}^2]^T$, i.e. the $m$-th \textit{power allocation vector}. Fortunately, the following proposition shows that $R_\text{LB}$ is actually a concave function of the joint power allocation vector $\bm\lambda \triangleq [\bm\lambda_1^T, \ldots, \bm\lambda_M^T]^T \in \mathbb{R}_{MN_\text{S} \times 1}$.

\begin{proposition}
The closed-form expression $R_\text{LB}$ is a concave function of the power allocation vector $\bm\lambda$.
\label{proposition2}
\end{proposition}

\begin{IEEEproof}
  The proof is provided in the Appendix \ref{AppC}.
\end{IEEEproof}

Aided with Proposition \ref{proposition2}, we therefore seek to solve the following convex optimization problem (P2-1) to obtain the global-optimal power allocation vector $\bm\lambda$:
\begin{equation}
\arraycolsep=1.0pt\def\arraystretch{1.3}
  \begin{array}{lcl}
  \text{(P2-1):    }& \displaystyle \max_{\bm\lambda \in \mathbb{R}_{MN_\text{S}\times 1}}               & \displaystyle R_\text{LB}(\bm\lambda) \,\, \text{s.t.} \,\, \mathbf{1}^T \bm\lambda = M N_\text{S}, \,\, \bm\lambda \succeq \mathbf{0},
  \end{array}
  \label{P2_1}
\end{equation}
where $\bm\lambda \succeq \mathbf{0}$ represents a non-negative vector $\bm\lambda$.

To solve (P2-1), we utilize the \textit{barrier method} to incorporate the non-negative constraint $\bm\lambda \succeq \mathbf{0}$ \cite{boyd2004convex}, i.e.
\begin{equation}
\arraycolsep=1.0pt\def\arraystretch{1.3}
  \begin{array}{cl}
  \displaystyle \max_{\bm\lambda \in \mathbb{R}_{MN_\text{S}\times 1}} & \displaystyle f_\text{B}(\bm\lambda) = R_\text{LB}(\bm\lambda) + \sum_{i=1}^{MN_\text{S}} \phi (\lambda_i) \,\, \text{s.t.} \,\, \mathbf{1}^T \bm\lambda = MN_\text{S} \\
  \end{array},
  \label{P2_1_Barrier}
\end{equation}
where $\lambda_i$ denotes the $i$-th element of $\bm\lambda$, and $\phi(u)$ is the logarithmic barrier function utilized to approximate the penalty of violating the non-negative constraint, i.e.
\begin{equation}
  \phi(u) = \begin{cases}
    \displaystyle \frac{1}{t_\text{B}} \ln(u), & u > 0, \\
    \displaystyle -\infty, & u \le 0,
  \end{cases}
  \label{PhiDefinition}
\end{equation}
where $t_\text{B}$ is used to scale the barrier function's penalty. In order to solve (\ref{P2_1_Barrier}), we formulate the gradient of the cost function in (\ref{P2_1_Barrier}) with respect to $\bm\lambda_m$ as follows:
\begin{equation}
  \nabla_{\bm\lambda_m}f_\text{B}(\bm\lambda) = \nabla_{\bm\lambda_m} R_\text{LB}(\bm\lambda) + \frac{1}{t_\text{B}}\mathbf{q}_m,
  \label{fB_lambda_m}
\end{equation}
where $\mathbf{q}_m \triangleq \left[\lambda_{m1}^{-1}, \ldots, \lambda_{mN_\text{S}}^{-1}\right]^T$. The expression of the gradient vector $\nabla_{\bm\lambda_m} R_\text{LB}(\bm\lambda)$ has been derived in (\ref{RLB_Lambda_m}). Based on (\ref{fB_lambda_m}), the gradient of $f_\text{B}(\bm\lambda)$ at $\bm\lambda$ is thus given by:
\begin{equation}
  \nabla_{\bm\lambda} f_\text{B}(\bm\lambda) = [\nabla_{\bm\lambda_1} f_\text{B}(\bm\lambda)^T, \ldots, \nabla_{\bm\lambda_M} f_\text{B}(\bm\lambda)^T]^T.
  \label{DBF_Gradient}
\end{equation}

To preserve the linear constraint $\mathbf{1}^T \bm\lambda = MN_\text{S}$, we therefore formulate the ascent direction as follows \cite{zeng2012linear}:
\begin{equation}
  \Delta \bm\lambda = \left(\mathbf{I}_{MN_\text{S}} - \frac{\mathbf{1} \cdot \mathbf{1}^T }{M N_\text{S}}\right) \nabla_{\bm\lambda}f_\text{B}(\bm\lambda),
  \label{DBF_SearchDirection}
\end{equation}
by which the gradient $\nabla_{\bm\lambda}f_\text{B}(\bm\lambda)$ is projected onto the linear space satisfying:
\begin{equation}
  \mathbf{1}^T \Delta \bm\lambda = 0.
\end{equation}

Finally, we summarize our digital precoder optimization algorithm in Algorithm \ref{alg:AlgorithmDBF}.

\begin{algorithm}[htb]
 \caption{Maximizing the SE Lower Bound Over the Baseband Power Allocation Vector}
 \label{alg:AlgorithmDBF}
 \begin{algorithmic}[1]
 \STATE \textit{Initialization}: Given a feasible initial solution $\bm\lambda^{(0)}$, $i=0$, halting criterion $\epsilon_\text{halt} > 0$ and the barrier coefficient $t_\text{B}$.

 \STATE \textit{Search direction}: Compute the gradient $\nabla_{\bm\lambda} f_\text{B}(\bm\lambda^{(i)})$ as (\ref{DBF_Gradient}) and the search direction $\Delta \bm\lambda^{(i)}$ as (\ref{DBF_SearchDirection}). \label{alg:AlgorithmDBF:SearchDirection}

 \STATE \textit{Gradient ascent}: Solve the following one-dimensional search problem via backtracking line search \cite{boyd2004convex}:
 \begin{equation}
   \eta^* = \arg \max_{\eta} f_\text{B}\left( \bm\lambda^{(i)} + \eta \cdot \Delta \bm\lambda^{(i)} \right).\nonumber
 \end{equation}

 \STATE \textit{Update}: Stop if $\eta^* \|\Delta\bm\lambda^{(i)}\|_2 \le \epsilon_\text{halt} \|\bm\lambda^{(i)}\|_2$, else let $\bm\lambda^{(i+1)} \leftarrow \bm\lambda^{(i)} + \eta^* \cdot \Delta \bm\lambda^{(i)}$, $i \leftarrow i + 1$ and then go to Step \ref{alg:AlgorithmDBF:SearchDirection}.

 \end{algorithmic}
\end{algorithm}

As the concavity of $R_\text{LB}(\bm\lambda)$ over $\bm\lambda$ has been verified, Algorithm \ref{alg:AlgorithmDBF} thus ensures convergence to a global optimal power allocation vector $\bm\lambda^*$. The optimal digital precoder $\mathcal{D}^* = \{\mathbf{D}_1^*, \ldots, \mathbf{D}_M^*\}$ is thus given as:
\begin{equation}
   \mathbf{D}_m^* = \displaystyle \text{diag}\left(\sqrt{\lambda_{m1}^*}, \sqrt{\lambda_{m2}^*}, \ldots, \sqrt{\lambda_{mN_\text{S}}^*}\right),\,\, 1 \le m \le M,
   \label{DBFOutputFormulation}
\end{equation}
where we have $\bm\lambda^* = [ (\bm\lambda^*_1)^T, \ldots, (\bm\lambda^*_M)^T ]^T$.

\subsection{Analog Precoder Design for SE Maximization}
Different from (P2-1), the optimization of (P3) is non-concave due to i) the non-convex constraint of $\mathbf{A} \in \mathscr{A}$, and ii) the non-concavity of $R_\text{LB}$ over $\mathbf{A}$, of which the latter can be again verified via a simple counterexample (e.g. $N_\text{R} = N_\text{T} = N_\text{K} = N_\text{M} = N_\text{RF} = M = 1$). In order to handle the non-convex constraint of $\mathbf{A} \in \mathscr{A}$, we propose to relax the problem (P3) into the following optimization (P3-1) with a convex $\ell_\infty$ constraint:
\begin{equation}
\arraycolsep=1.0pt\def\arraystretch{1.3}
  \begin{array}{lcl}
  \text{(P3-1):    }& \displaystyle \max_{\mathbf{a} \in \mathbb{C}_{N_\text{T}\times 1}}               & \displaystyle R_\text{LB}(\mathbf{a}) \,\, \text{s.t.} \,\, \left\|\mathbf{a}\right\|_\infty \le 1/\sqrt{N_\text{K}},
  \end{array}
  \label{P3_1}
\end{equation}
where $\mathbf{a} \in \mathbb{C}_{N_\text{T} \times 1}$ denotes the diagonal elements of $\mathbf{A}$, i.e. $\mathbf{a} = \text{diag}(\mathbf{A})$. Note that the original feasible set $\mathscr{A}$ is a subset of the new feasible set in (P3-1), i.e.
\begin{equation}
  \left\{ \mathbf{a} \in \mathbb{C}_{N_\text{T} \times 1}:  a_i = \frac{\exp\left(j \theta\right)}{\sqrt{N_\text{K}}}, \,\, 1 \le i \le N_\text{T} \right\} \subset \left\{\mathbf{a} \in \mathbb{C}_{N_\text{T} \times 1}: \left\|\mathbf{a}\right\|_\infty \le \frac{1}{\sqrt{N_\text{K}}}\right\},
\end{equation}
while the feasible set of (P3-1) is also convex due to the convexity of $\ell_\infty$ norm.

In order to deal with the non-differentiable $\ell_\infty$ constraint in (P3-1), similar to \cite{han2014binary}, we exploit the $\ell_p$ approximation with a large $p$. Since
\begin{equation}
  \lim_{p \rightarrow \infty} \left\| \mathbf{a} \right\|_p = \left\|\mathbf{a} \right\|_\infty,
\end{equation}
the value of $p$ should thus be gradually increased during the optimization process. Moreover, again we exploit the logarithmic barrier function to approximate the penalty of violating the $\ell_p$ constraint, which leads to the following optimization problem:
\begin{equation}
\arraycolsep=1.0pt\def\arraystretch{1.3}
  \begin{array}{cl}
  \displaystyle \max_{\mathbf{a} \in \mathbb{C}_{N_\text{T} \times 1}}  & \displaystyle g_\text{B}(\mathbf{a}, p) = R_\text{LB}(\mathbf{a}) + \phi\left(\frac{1}{\sqrt{N_\text{K}}} - \left\|\mathbf{a}\right\|_p\right),
  \end{array}
  \label{P3_2}
\end{equation}
where the barrier function $\phi(u)$ has been defined in (\ref{PhiDefinition}). To solve (\ref{P3_2}) via a gradient method, we formulate the gradient of the cost function $g_\text{B}(\mathbf{a}, p)$ over $\mathbf{a}$ as follows:
\begin{equation}
  \nabla_{\mathbf{a}} g_\text{B}(\mathbf{a}, p) = \nabla_\mathbf{a} R_\text{LB}(\mathbf{a}) -  \frac{\left\|\mathbf{a}\right\|_p^{1-p} \mathbf{p}_\text{a} }{2 t_\text{B} \left( N_\text{K}^{-1/2} - \left\|\mathbf{a}\right\|_p \right)},
  \label{gB_a}
\end{equation}
where $\mathbf{p}_\text{a} \in \mathbb{C}_{N_\text{T} \times 1}$ is given as:
\begin{equation}
  \mathbf{p}_\text{a} = \left[ a_1 \cdot \left|a_1\right|^{p-2}, a_2 \cdot \left|a_2\right|^{p-2}, \ldots, a_{N_\text{T}} \cdot \left|a_{N_\text{T}}\right|^{p-2} \right]^T.
\end{equation}

\begin{figure*}
\small
    \begin{equation}
      \nabla_\mathbf{a} R_\text{LB}(\mathbf{a}) = \frac{\rho \log_2 e}{M N_\text{S}} \sum_{n=1}^M \frac{\sum_{t=1}^M \left|\bm\Sigma_n + \bm\Sigma_t\right|^{-1} \text{diag}\left[\mathbf{H}^H \left(\bm\Sigma_n + \bm\Sigma_t\right)^{-1} \mathbf{HA} \left(\mathbf{C}_n \bm\Lambda_n \mathbf{C}_n^H + \mathbf{C}_t \bm\Lambda_t \mathbf{C}_t^H \right)\right] }{\sum_{t'=1}^M \left|\bm\Sigma_n + \bm\Sigma_{t'}\right|^{-1}}.
      \label{RLB_a}
    \end{equation}
\hrulefill
\end{figure*}

Moreover, the gradient $\nabla_\mathbf{a} R_\text{LB}(\mathbf{a})$ is given in (\ref{RLB_a}), where $\bm\Lambda_n = \text{diag}(\bm\lambda_n)$ for $n = 1, 2, \ldots, M$. By using $\nabla_\mathbf{a} g_\text{B}(\mathbf{a}, p)$ as the search direction, i.e. $\Delta \mathbf{a} = \nabla_\mathbf{a} g_\text{B}(\mathbf{a}, p)$, we thus present our proposed algorithm for the analog precoder design in Algorithm \ref{alg:AlgorithmABF}.

\begin{algorithm}[htb]
 \caption{Maximizing the SE Lower Bound Over the Analog Precoder}
 \label{alg:AlgorithmABF}
 \begin{algorithmic}[1]
 \STATE \textit{Initialization}: Given a feasible initial solution $\mathbf{a}^{(0)}$, $p > 0$, $\Delta p > 0$, $p_\text{max} > 0$, $i=0$, halting criterion $\epsilon_\text{halt} > 0$ and the barrier coefficient $t_\text{B}$.\label{alg:AlgorithmABF:Initialization}

 \STATE \textit{Search direction}: Compute the search direction $\Delta \mathbf{a}^{(i)} = \nabla_\mathbf{a} g_\text{B}(\mathbf{a}^{(i)}, p)$ with $\nabla_\mathbf{a} g_\text{B}(\mathbf{a}^{(i)}, p)$ given in (\ref{gB_a}).\label{alg:AlgorithmABF:SearchDirection}

 \STATE \textit{Gradient ascent}: Solve the following one-dimensional search problem via backtracking line search:
 \begin{equation}
   \eta^* = \arg\max_\eta g_\text{B}(\mathbf{a}^{(i)} + \eta \cdot \Delta \mathbf{a}^{(i)}, p).\nonumber
 \end{equation}

 \STATE \textit{Update}: Go to Step \ref{alg:AlgorithmABF:Iteration} if $\eta^* \|\Delta \mathbf{a}^{(i)}\|_2 \le \epsilon_\text{halt} \|\mathbf{a}^{(i)}\|_2$, else let $\mathbf{a}^{(i+1)} \leftarrow \mathbf{a}^{(i)} + \eta^* \cdot \Delta \mathbf{a}^{(i)}$, $i \leftarrow i+1$ and then go to Step \ref{alg:AlgorithmABF:SearchDirection}.

 \STATE \textit{Iteration}: Go Step \ref{alg:AlgorithmABF:Output} if $p \ge p_\text{max}$, else let $p \leftarrow p + \Delta p$ and then go to Step \ref{alg:AlgorithmABF:SearchDirection}. \label{alg:AlgorithmABF:Iteration}

 \STATE \textit{Output}: The optimized analog precoder's diagonal elements are thus given by:
 \begin{equation}
    \mathbf{a}^* = \frac{1}{\sqrt{N_\text{K}}} \exp\left[j \text{angle}\left(\mathbf{a}^{(i)}\right)\right],
 \end{equation}
 where $\text{angle}(\cdot)$ represents the element-wise phase function.\label{alg:AlgorithmABF:Output}
 \end{algorithmic}
\end{algorithm}

Due to the non-concavity of $R_\text{LB}$ over $\mathbf{a}$ and the convex $\ell_\infty$-norm relaxation, Algorithm \ref{alg:AlgorithmABF} thus ensures convergence to a local maximum of $\mathbf{a}$.

\subsection{Proposed Two-Step Algorithm for Hybrid Precoder Design}
By combining Algorithm \ref{alg:AlgorithmDBF} and \ref{alg:AlgorithmABF}, we therefore develop our proposed two-step algorithm for the hybrid precoder design in Algorithm \ref{alg:AlgorithmHybrid}, where the digital precoder $\mathcal{D}$ and the analog precoder $\mathbf{A}$ are optimized \textit{iteratively}.

\begin{algorithm}[htb]
 \caption{Two-Step Algorithm for Hybrid Precoder Design}
 \label{alg:AlgorithmHybrid}
 \begin{algorithmic}[1]
 \STATE \textit{Initialization}: Given initial solutions $\mathbf{a}^{(0)}$ and $\bm\lambda^{(0)}$. Set iteration index to $i=0$.

 \STATE \textit{Optimize the digital precoder}: Based on $\mathbf{a}^{(i)}$, optimize the digital precoder via Algorithm \ref{alg:AlgorithmDBF} and (\ref{DBFOutputFormulation}), which yields $\bm\lambda^{(i+1)}$. \label{alg:AlgorithmHybrid:DBF}

 \STATE \textit{Optimize the analog precoder}: Based on $\bm\lambda^{(i+1)}$, optimize the analog precoder via Algorithm \ref{alg:AlgorithmABF}, which yields $\mathbf{a}^{(i+1)}$. \label{alg:AlgorithmHybrid:ABF}

 \STATE Let $i \leftarrow i + 1$. Go to Step \ref{alg:AlgorithmHybrid:DBF} until convergence.
 \end{algorithmic}
\end{algorithm}

As Algorithm \ref{alg:AlgorithmHybrid} only ensures convergence to a local maximum (since Algorithm \ref{alg:AlgorithmABF} only ensures local convergence), the optimization results of Algorithm \ref{alg:AlgorithmHybrid} are thus affected by the initialization of $\mathbf{a}^{(0)}$ and $\bm\lambda^{(0)}$. However, as we will show in the next subsection, the optimized cost function is relatively insensitive to the specific selection of initial points.

\subsection{Convergence of the Proposed Two-Step Algorithm}
In this subsection, several examples will be provided to confirm the convergence of the proposed Algorithm \ref{alg:AlgorithmHybrid} in conjunction with various initial points. Note that the simulation parameters are configured according to Table \ref{TABLESimu} with an SNR value of $5$ dB. The initial solutions of Algorithm \ref{alg:AlgorithmHybrid} are designed as follows:
\begin{equation}
\arraycolsep=1.0pt\def\arraystretch{2.2}
  \begin{array}{rcl}
  \displaystyle \bm\lambda^{(0)} &=& \displaystyle \frac{MN_\text{S}}{\sum_{n=1}^{MN_\text{S}} \lambda_n^{(0)}} \cdot \left[\lambda_1^{(0)}, \lambda_2^{(0)}, \ldots, \lambda_{MN_\text{S}}^{(0)}\right]^T, \\
  \displaystyle \mathbf{a}^{(0)} &=& \displaystyle \left[ \frac{\exp(j\theta_1)}{\sqrt{N_\text{K}}}, \frac{\exp(j\theta_2)}{\sqrt{N_\text{K}}}, \ldots, \frac{\exp(j\theta_{N_\text{T}})}{\sqrt{N_\text{K}}} \right]^T,
  \end{array}
\end{equation}
where $\lambda_n^{(0)}$ and $\theta_m$ ($1 \le n \le MN_\text{S}$, $1 \le m \le N_\text{T}$) are i.i.d. random variables subject to a uniform distribution over $[0, 1]$ and $[-\pi, \pi]$, respectively.

\begin{figure}
\center{\includegraphics[width=0.75\linewidth]{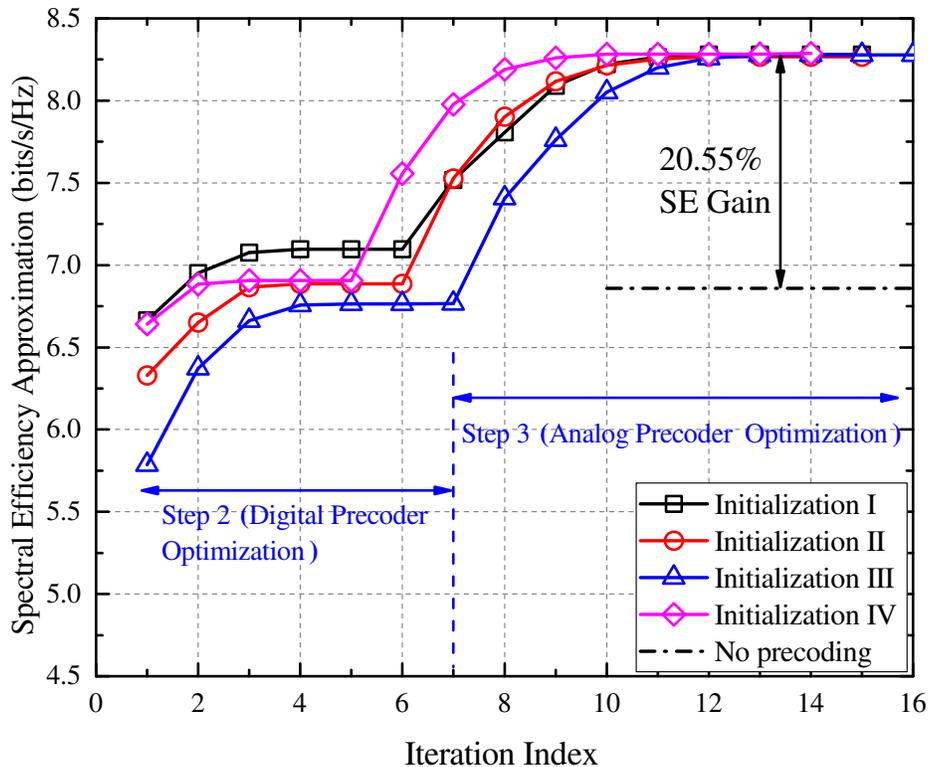}}
\caption{Typical evolution yielded by Algorithm \ref{alg:AlgorithmHybrid} with $4$ independently generated initial solutions. The simulation parameters are specified according to Table \ref{TABLESimu} with $p = 32$, $\Delta p = 10$, $p_\text{max} = 64$, $\epsilon_\text{halt} = 10^{-3}$ and $t_\text{B} = 64$.}
\label{Fig_TypicalEvo}
\end{figure}

The evolution of the proposed Algorithm \ref{alg:AlgorithmHybrid} in conjunction with $4$ independently generated initial points is therefore presented in Fig.\ref{Fig_TypicalEvo}. Note that here we use the closed-form SE approximation, i.e. $[R_\text{LB} - N_\text{R}(1-\log_2 e)]$ as the performance metric. The performance yielded without precoding is also depicted in Fig.\ref{Fig_TypicalEvo}. It can thus be that Algorithm \ref{alg:AlgorithmHybrid} converges to almost the same cost function value for the various initial points, which outperforms the SE without precoding by approximately $20.55 \%$. Note that the evolution of the cost function exhibits a staircase shape with each stair associated with either Step \ref{alg:AlgorithmHybrid:DBF} or Step \ref{alg:AlgorithmHybrid:ABF} of Algorithm \ref{alg:AlgorithmHybrid}. Besides, it is also observed that it only takes less than $20$ iterations for Algorithm \ref{alg:AlgorithmHybrid} to converge, which substantiates the low complexity advantage of the proposed algorithm.

\begin{figure}
\center{\includegraphics[width=0.75\linewidth]{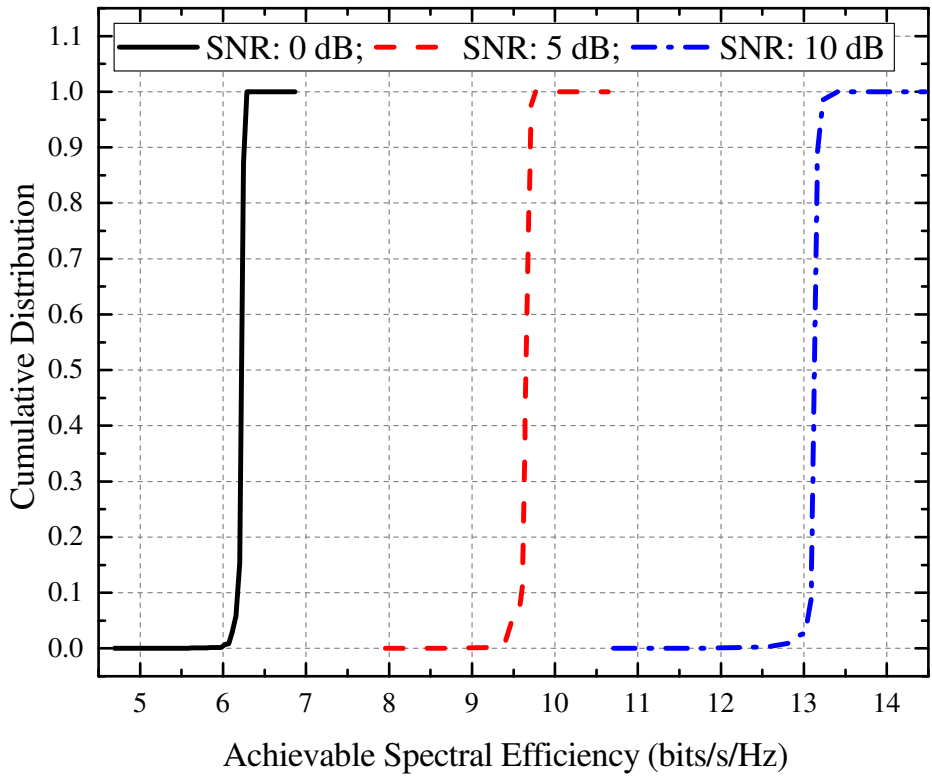}}
\caption{Cumulative distribution of the achievable SE yielded by various initial points. The simulation parameters are configured in accordance to Fig.\ref{Fig_TypicalEvo}.}
\label{Fig_ConvergenceProof}
\end{figure}

In order to provide a more intuitive demonstration, in Fig.\ref{Fig_ConvergenceProof} we depict the cumulative distribution of the achievable SE yielded by $10,000$ randomly generated initial points. The simulation parameters are configured in accordance to Fig.\ref{Fig_TypicalEvo} in conjunction with SNR $\in \{0, 5, 10\}$ dB. Based on the steeply ascending shape of the curves depicted in Fig.\ref{Fig_ConvergenceProof}, it can thus be concluded that the proposed Algorithm \ref{alg:AlgorithmHybrid} ensures convergence to almost the same cost function for all the initial points, which therefore confirms the near global optimality achieved by the proposed algorithm.

\subsection{Complexity Analysis}
We now provide analysis on the computational complexity of the proposed Algorithm \ref{alg:AlgorithmHybrid}. We commence by quantifying the complexity order of Algorithm \ref{alg:AlgorithmDBF} and Algorithm \ref{alg:AlgorithmABF} as follows.

\textit{(1) Complexity of Algorithm \ref{alg:AlgorithmDBF}}: As can be seen from Algorithm \ref{alg:AlgorithmDBF}, the computational complexity is primarily consumed by the gradient calculation, which involves i) calculating the $M^2$ matrices' inversions $(\bm\Sigma_n + \bm\Sigma_m)^{-1}$ for $1 \le n, m \le M$, and ii) calculating the $M^2$ matrix multiplications $\mathbf{C}_m^H\mathbf{A}^H \mathbf{H}^H (\bm\Sigma_n + \bm\Sigma_m)^{-1}\mathbf{HAC}_m$. Therefore the complexity order of Algorithm \ref{alg:AlgorithmDBF} for each iteration is:
\begin{equation}
  \mathcal{O}\left[M^2 \left(N_\text{R}^3 + 2N_\text{RF} N_\text{R}^2\right)\right].
\end{equation}

\textit{(2) Complexity of Algorithm \ref{alg:AlgorithmABF}}: Similar to Algorithm \ref{alg:AlgorithmDBF}, the complexity of Algorithm \ref{alg:AlgorithmABF} is also mainly consumed by the gradient calculation, which involves calculating the $\ell_p$ norm $\|\mathbf{a}\|_p$ as well as the following matrices ($1 \le n, t \le M$):
\begin{equation}
  \mathbf{H}^H \left(\bm\Sigma_n + \bm\Sigma_t\right)^{-1} \mathbf{HA} \left( \mathbf{C}_n\bm\Lambda_n\mathbf{C}_n^H + \mathbf{C}_t \bm\Lambda_t \mathbf{C}_t^H \right).
\end{equation}

Therefore the complexity order of Algorithm \ref{alg:AlgorithmABF} for each iteration is:
\begin{equation}
  \mathcal{O}\left[M^2 \left(N_\text{R}^3 + N_\text{T}N_\text{R}^2 + N_\text{R} N_\text{T}^2 \right) + p N_\text{T}\right].
\end{equation}

Finally, by preserving the dominant terms, the overall complexity order of Algorithm \ref{alg:AlgorithmHybrid} can be expressed as follows:
\begin{equation}
  \mathcal{O}\left[ M^2 \left( N_\text{R}^3 + 2 N_\text{RF} N_\text{R}^2 \right) + N_\text{p} \cdot M^2 \left( N_\text{R}^3 + N_\text{T}N_\text{R}^2 + N_\text{R} N_\text{T}^2 \right) + N_\text{T} \sum_{n=1}^{N_\text{p}} p^{(n)} \right],
  \label{AlgorithmHybridComOrder}
\end{equation}
where $N_\text{p} = \left\lceil \left( p_\text{max} - p \right) / \Delta p \right\rceil$ and $p^{(n)} = p + (n-1) \Delta p$, with $p$, $\Delta p$ and $p_\text{max}$ specified in Step \ref{alg:AlgorithmABF:Initialization} of Algorithm \ref{alg:AlgorithmABF}.

It is worth noting that, the polynomial complexity order in (\ref{AlgorithmHybridComOrder}) is mainly achieved thanks to the application of the closed-form cost function $R_\text{LB}$. Otherwise the complexity order would be orders of magnitude higher due to the prohibitive complexity for calculating the true SE expression $R(\mathbf{H}, \mathcal{D}, \mathbf{A})$.

\subsection{Optimization of System Parameters}
In this subsection we discuss the optimized selection of the system parameters, i.e $N_\text{T}$, $N_\text{R}$, $N_\text{K}, N_\text{M}$ and $N_\text{RF}$. As the antennas and RF chains are usually hardware resources that are invariant from a practical point of view, we therefore focus on the selection of $(N_\text{K}, N_\text{M})$.

Note that in the proposed scheme $N_\text{T} = N_\text{K} N_\text{M}$ always holds. On the one hand, increasing $N_\text{K}$ leads to reducing $N_\text{M}$, which consequently reduces the potential multiplexing gain provided by GenSM, since $M = 2^{\lfloor \log_2 \binom{N_\text{M}}{N_\text{RF}} \rfloor}$. On the other hand, increasing $N_\text{K}$ also leads to a larger antenna group and therefore enhances the possible array gain provided by incorporating the ananlog precoder $\mathbf{A}$. Therefore the pair $(N_\text{K}, N_\text{M})$ is essential for achieving a scalable tradeoff between multiplexing gain and array gain. As $(N_\text{K}, N_\text{M})$ cannot be altered for every channel realization $\mathbf{H}$, we thus seek to optimize the parameters for maximizing the \textit{average SE}, i.e.
\begin{equation}
\arraycolsep=1.0pt\def\arraystretch{1.3}
  \begin{array}{rcl}
  \left(N_\text{K}^*, N_\text{M}^* \right) &=& \displaystyle \arg\max_{(N_\text{K}, N_\text{M})} E_\mathbf{H}\left\{R_\text{LB}\left[\mathbf{H}, \mathcal{D}^*(\mathbf{H}, N_\text{K}, N_\text{M}), \mathbf{A}^*(\mathbf{H}, N_\text{K}, N_\text{M}) \right]\right\}, \\
  && \displaystyle \text{s.t.} \,\, N_\text{K} N_\text{M} = N_\text{T},
  \end{array}
  \label{ParameterOptimization}
\end{equation}
where $\mathcal{D}^*(\mathbf{H}, N_\text{K}, N_\text{M})$ and $\mathbf{A}^*(\mathbf{H}, N_\text{K}, N_\text{M})$ denote the hybrid precoder designed by Algorithm \ref{alg:AlgorithmHybrid}, when $\mathbf{H}$, $N_\text{K}$ and $N_\text{M}$ are given. Note that we use $R_\text{LB}$ instead of $R$ as the cost function in (\ref{ParameterOptimization}) so that a lower complexity can be obtained.

Using (\ref{ParameterOptimization}) as the design guideline, we present the optimal $(N_\text{K}, N_\text{M})$ pairs as a function of various configuration parameters in Table \ref{TABLEParameterOptimization}. As it can be seen from Table \ref{TABLEParameterOptimization}, the optimal value of $N_\text{M}$ is shown to increase with the increase of $N_\text{R}$ or SNR, i.e. a larger $N_\text{M}$ should be invoked, when the receiver is in a sufficiently good condition (either a larger $N_\text{R}$ or a higher SNR value) to harness the SMX gain provided by GenSM. Otherwise, when a lower SNR or a smaller $N_\text{R}$ value is invoked, $N_\text{M}$ should be reduced to enhance the array gain provided by analog beamforming.

\begin{table}
\caption{Optimal $(N_\text{K}, N_\text{M})$ Pairs as a Function of Various System Parameters}
\newcommand{\tabincell}[2]{\begin{tabular}{@{}#1@{}}#2\end{tabular}}
\centering
\renewcommand\arraystretch{1.2}
\begin{tabular}{c|c|c|c|c|c}
\hline\hline
 &  & \multicolumn{4}{c}{SNR (dB)} \\\hline
$N_\text{RF}$ & $N_\text{T} \times N_\text{R}$ & -5 & 0 & 5 & 10 \\\hline
\multirow{2}{*}{1} & $8 \times 4$ & $(8, 1)$ & $(8, 1)$ & $(4, 2)$ & $(1, 8)$ \\\cline{2-6}
& $8 \times 8$ & $(8, 1)$ & $(2, 4)$ & $(1, 8)$ & $(1, 8)$ \\\hline
\multirow{2}{*}{2} & $8 \times 6$ & $(4, 2)$ & $(4, 2)$ & $(2, 4)$ & $(1, 8)$ \\\cline{2-6}
& $8 \times 8$ & $(4, 2)$ & $(4, 2)$ & $(2, 4)$ & $(1, 8)$\\\hline
\hline
\end{tabular}
\label{TABLEParameterOptimization}
\end{table}


\textit{Remark}: It is worth noting that our proposed scheme degenerates to the conventional sub-connected hybrid precoding schemes, when $(N_\text{K}, N_\text{M}) = (N_\text{T}/N_\text{RF}, N_\text{RF})$. Hence the proposed scheme has the potential to even outperform the conventional schemes in terms of achievable SE. As a matter of fact, the solution to (\ref{ParameterOptimization}) is the essential reason for the performance improvements achieved by the proposed scheme, as the conventional schemes can be conceived as special cases of the proposed GenSM-aided mmWave MIMO scheme. The performance improvements will be substantiated in the following sections.

\section{Simulation Results}
In this section we present the simulated SE performance yielded by various schemes. Note that the achievable SE performance of the proposed scheme is given by the \textit{true SE expression} $R(\mathbf{H}, \mathcal{D}, \mathbf{A})$ averaged over $1, 000$ random channel realizations. The simulation parameters (e.g. $p$, $\Delta p$, $p_\text{max}$, etc.) are specified as in Fig.\ref{Fig_TypicalEvo} and Table \ref{TABLESimu}, unless mentioned otherwise.

More specifically, the achievable SE performance yielded by the following $5$ schemes are presented for performance comparison:
\begin{itemize}
  \item \textit{O-GenSM-MIMO}: The proposed GenSM-aided mmWave MIMO scheme with hybrid precoder optimized according to Algorithm \ref{alg:AlgorithmHybrid}. The performance of O-GenSM-MIMO is associated with $(N_\text{T}, N_\text{R}, N_\text{RF})$, while $(N_\text{K}, N_\text{M})$ are selected according to (\ref{ParameterOptimization}).

  \item \textit{NO-GenSM-MIMO}: The proposed scheme without optimization. The system parameters of NO-GenSM-MIMO are configured in accordance to the corresponding O-GenSM-MIMO counterpart.

  \item \textit{WP-MIMO}: Waterfilling-precoded MIMO scheme \cite{perez2010MIMO}. Note that, in conventional WP-MIMOs, $N_\text{S}$ is usually set as $N_\text{S} = N_\text{T}$ to fully exploit the spatial multiplexing gain. In order to maintain fairness from an RF-chain-limited point of view, we thus assume that $N_\text{S} = N_\text{RF}$ also holds for WP-MIMOs.

  \item \textit{SIC-SC-MIMO}: Hybrid precoding scheme for the sub-connected (SC) mmWave MIMO using successive interference cancellation (SIC) method, which is proposed by \cite{gao2016energy}.

  \item \textit{S-Sparse-MIMO}: The classic spatially sparse hybrid precoded mmWave MIMO proposed by \cite{ayach2014spatially}. Note that S-Sparse-MIMO exploits a full-connected hybrid precoder structure, which requires more hardware complexity than the sub-connected structure exploited by our scheme.
\end{itemize}

\begin{figure}
\center{\includegraphics[width=0.75\linewidth]{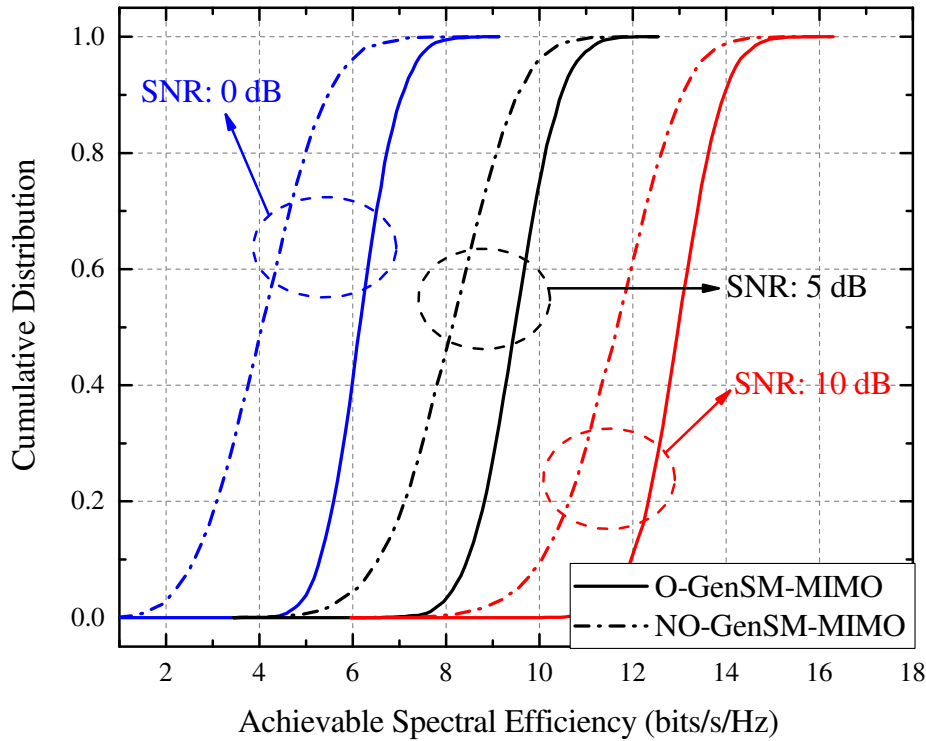}}
\caption{Cumulative distribution of the achievable SE yielded by O-GenSM-MIMO and NO-GenSM-MIMO with various channel realizations. The parameters are $N_\text{T}=8$, $N_\text{R}=8$ and $N_\text{RF}=2$, while $(N_\text{K}, N_\text{M})$ are determined by solving (\ref{ParameterOptimization}).}
\label{Fig_PerformanceCDF}
\end{figure}

We commence by showing the cumulative distribution of the achievable SE yielded by O-GenSM-MIMO and NO-GenSM-MIMO with various channel realizations in Fig.\ref{Fig_PerformanceCDF}. As it can be seen from the figure, aided with the proposed optimization algorithm, the SE achieved by O-GenSM-MIMO is capable of significantly outperforming the SE achieved by NO-GenSM-MIMO, which substantiates the efficacy of the proposed Algorithm \ref{alg:AlgorithmHybrid}. Moreover, it can be also observed that the cumulative distribution of O-GenSM-MIMO is even steeper than that of NO-GenSM-MIMO, which indicates that the channel variation has less impacts on the performance of O-GenSM-MIMO than NO-GenSM-MIMO, i.e. O-GenSM-MIMO is more robust under the channel fading.

\begin{figure}
\center{\includegraphics[width=0.75\linewidth]{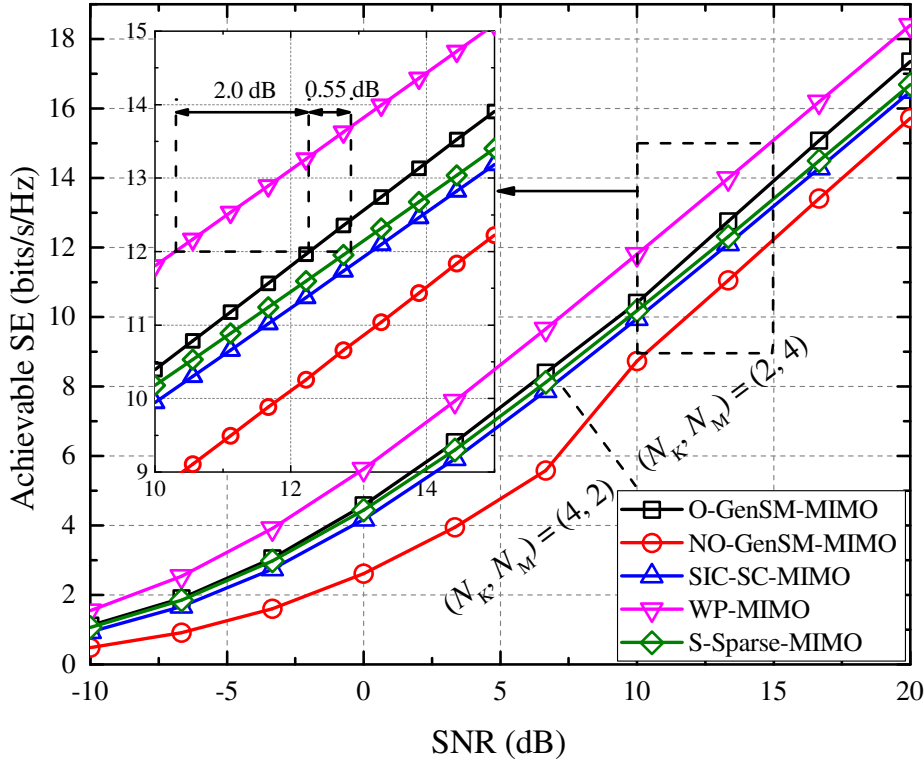}}
\caption{Achievable SE performance yielded by various schemes with $N_\text{T} = 8$, $N_\text{R} = 4$ and $N_\text{RF} = 2$. The parameters $(N_\text{K}, N_\text{M})$ of the proposed scheme are designed by solving (\ref{ParameterOptimization}).}
\label{Fig_Performance8x4}
\end{figure}

In Fig.\ref{Fig_Performance8x4}, the average SE performance yielded by various schemes with $(N_\text{T}, N_\text{R}, N_\text{RF}) = (8, 4, 2)$ are presented. For the proposed schemes, i.e. O-GenSM-MIMO and NO-GenSM-MIMO, it can be seen that $(N_\text{K}, N_\text{M}) = (4, 2)$ and $(N_\text{K}, N_\text{M}) = (2, 4)$ are respectively selected, when SNR $< 7.5$ dB and SNR $> 7.5$ dB. It is also observed that a significant SE improvement is achieved by O-GenSM-MIMO compared to NO-GenSM-MIMO, which substantiates the efficacy of the proposed hybrid precoder design in Algorithm \ref{alg:AlgorithmHybrid}. Furthermore, by comparing O-GenSM-MIMO to other state-of-the-art mmWave schemes, it is seen that our proposed scheme maintains a superior SE performance over the SIC-SC-MIMO scheme of \cite{gao2016energy} for the entire SNR range considered, and our scheme also outperforms the S-Sparse-MIMO scheme of \cite{ayach2014spatially} when the SNR is higher than $0$ dB. As predicted by the remarks in the last section, such performance improvement is guaranteed because our proposed scheme maintains a more generalized hybrid precoding paradigm, and the configuration parameters $(N_\text{K}, N_\text{M})$ are also optimized in terms of SE maximization, as in (\ref{ParameterOptimization}). Finally, with a target throughput of $12$ bits/s/Hz, our scheme outperforms the S-Sparse-MIMO scheme by about $0.55$ dB, while the WP-MIMO scheme outperforms the proposed scheme by approximately $2.0$ dB.

\begin{figure}
\center{\includegraphics[width=0.75\linewidth]{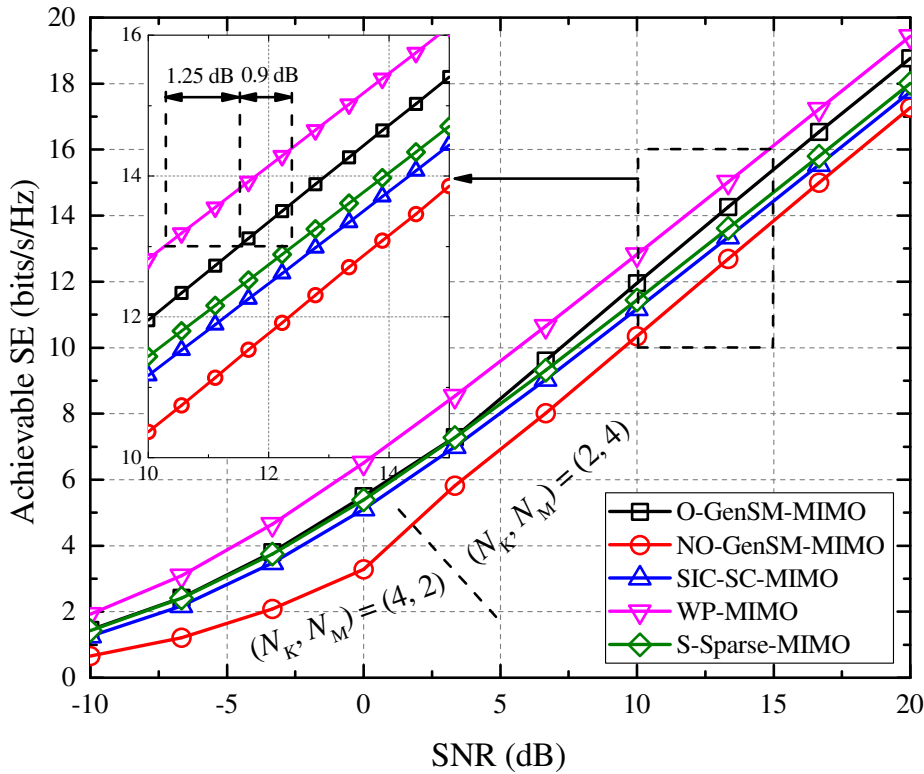}}
\caption{Achievable SE performance yielded by various schemes with $N_\text{T} = 8$, $N_\text{R} = 6$ and $N_\text{RF} = 2$. The parameters $(N_\text{K}, N_\text{M})$ of the proposed scheme are designed by solving (\ref{ParameterOptimization}).}
\label{Fig_Performance8x6}
\end{figure}

To explore the impact of $N_\text{R}$, we increase the $N_\text{R} = 4$ scenario in Fig.\ref{Fig_Performance8x4} to $N_\text{R} = 6$ and present Fig.\ref{Fig_Performance8x6}. It is seen that, with a higher number of RAs, $(N_\text{K}, N_\text{M}) = (4, 2)$ and $(N_\text{K}, N_\text{M}) = (2, 4)$ are respectively selected, when SNR $< 2.5$ dB and SNR $> 2.5$ dB, i.e. the ``SNR switching threshold'' is lower than the case with $N_\text{R}=4$. By comparing against other mmWave MIMO schemes, it is observed that the proposed scheme maintains a superior SE performance over S-Sparse-MIMO when SNR $> 2.5$ dB, and outperforms SIC-SC-MIMO over the entire SNR range under consideration. Finally, with a target throughput of $13$ bits/s/Hz, the proposed scheme outperforms S-Sparse-MIMO by approximately $0.9$ dB, and is outperformed by WP-MIMO with a $1.25$ dB performance gap.

\begin{figure}
\center{\includegraphics[width=0.75\linewidth]{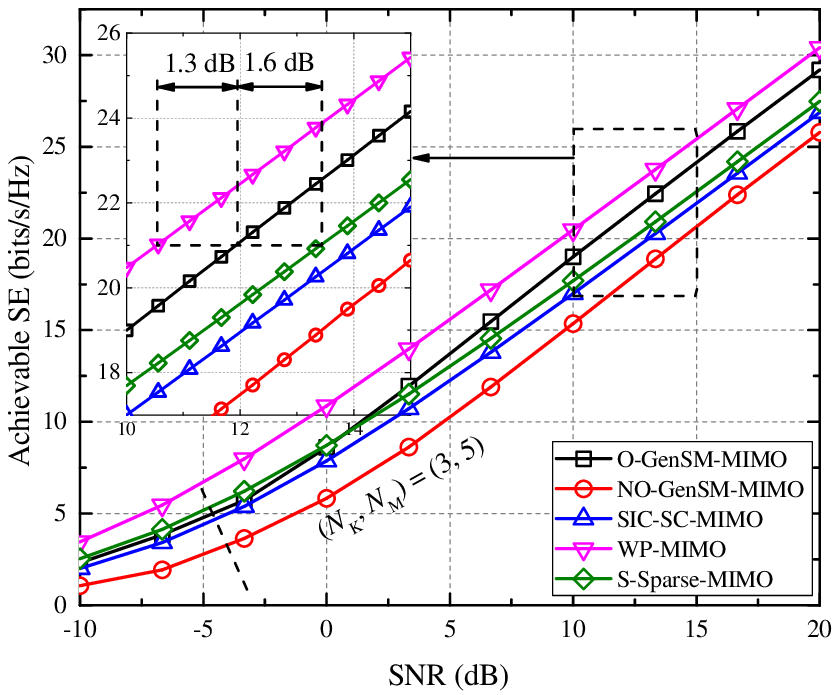}}
\caption{Achievable SE performance yielded by various schemes with $N_\text{T} = 15$, $N_\text{R} = 10$ and $N_\text{RF} = 2$. The parameters $(N_\text{K}, N_\text{M})$ of the proposed scheme are designed by solving (\ref{ParameterOptimization}).}
\label{Fig_Performance15x10}
\end{figure}

Finally, we increase the $N_\text{RF}=2$ cases to the case with $N_\text{RF}=3$ and present Fig.\ref{Fig_Performance15x10}, where a $15 \times 10$ mmWave MIMO is considered in conjunction with $3$ RF chains. It is seen that $(N_\text{K}, N_\text{M}) = (5, 3)$ and $(N_\text{K}, N_\text{M}) = (3, 5)$ are utilized, when SNR $< -7.5$ dB and SNR $> -7.5$ dB, respectively. Moreover, it is also observed that the proposed scheme maintains a higher SE performance than SIC-SC-MIMO for the entire SNR range considered. With  a target throughput of $21$ bits/s/Hz, it is readily seen that the proposed scheme outperforms S-Sparse-MIMO by approximately $1.6$ dB, and is outperformed by WP-MIMO with a $1.3$ dB performance gap.

To sum up, it can be observed from the simulation results that our proposed scheme is capable of outperforming the classic sub-connected mmWave scheme, i.e. SIC-SC-MIMO, for a wide range of SNR. The proposed scheme also outperforms the S-Sparse-MIMO scheme when a not-so-low SNR value is imposed. Note that such performance improvement is achieved with an even lower complexity level, consider that our scheme is sub-connected while S-Sparse-MIMO is full-connected. Finally, the proposed scheme remains sub-optimal with a very smaller SE gap, when compared against the optimal WP-MIMO scheme.

\section{Conclusion}
In this paper, we proposed a novel GenSM-aided mmWave MIMO scheme with a hybrid analog and digital precoding structure. A closed-form expression was proposed to quantify the achievable SE of the proposed scheme. Using the proposed expression as a low-complexity cost function, we proposed a new two-step algorithm to design the hybrid precoder with respect to SE maximization. More specifically, the proposed algorithm utilized the concavity of the cost function over the digital power allocation vector, and used a convex $\ell_\infty$ relaxation to handle the non-convex constraint imposed by the analog precoder. Finally, numerical simulation results not only demonstrated the convergence and efficacy of the proposed algorithm, but also substantiated the superior SE performance achieved by the proposed scheme against state-of-the-art mmWave precoding schemes.

\appendices
\section{Proof of Theorem \ref{theorem1}}\label{AppA}
\begin{IEEEproof}
The MI term in (\ref{MI0}) can be decomposed as follows \cite{cover2006elements}:
\begin{equation}
  I(\mathbf{y}; \mathbf{x}, m) = I(\mathbf{y}; \mathbf{x} \vert m) + I(\mathbf{y}; m),
  \label{RTotal}
\end{equation}
where $I(\mathbf{y}; \mathbf{x} \vert m)$ represents the average mutual information conditioned on a given $m$, which can be readily formulated using Shannon's continuous-input continuous-output memoryless channel's (CCMC) capacity \cite{za2014mianalysis}, i.e.
\begin{equation}
  I(\mathbf{y}; \mathbf{x} \vert m) = \displaystyle \frac{1}{M}\sum_{m=1}^M \log_2\left(\left|\frac{1}{\sigma_\text{N}^2} \bm\Sigma_m \right|\right),
  \label{RSym}
\end{equation}
where $\bm\Sigma_m$ is defined by (\ref{SigmanDef}). Moreover, the MI term $I(\mathbf{y}; m)$ represents the mutual information conveyed via the antenna domain, of which the expression is given by:
\begin{equation}
  I(\mathbf{y}; m) = \displaystyle \frac{1}{M} \sum_{n=1}^M \displaystyle \int \mathcal{P}(\mathbf{y} \vert n ) \log_2 \left[ \frac{\mathcal{P}(\mathbf{y} \vert n)}{\frac{1}{M} \sum_{t=1}^M \mathcal{P}(\mathbf{y} \vert t)} \right] \text{d} \mathbf{y},
  \label{RSpcForm1}
\end{equation}
where the likelihood function is given by:
\begin{equation}
  \mathcal{P}(\mathbf{y} \vert n) = \mathcal{CN}(\mathbf{y}; \mathbf{0}, \bm\Sigma_n).
\end{equation}

Since $I(\mathbf{y}; m)$ cannot be expressed in a closed form due to the discrete input $m$, we therefore derive a lower bound for $I(\mathbf{y}; m)$ as follows:
\begin{equation}
  \displaystyle I(\mathbf{y}; m) = \frac{1}{M} \sum_{n=1}^M \left\{ \int \mathcal{P}(\mathbf{y} \vert n) \log_2 \mathcal{P}(\mathbf{y} \vert n) \text{d} \mathbf{y}- \int \mathcal{P}(\mathbf{y}\vert n) \log_2 \left[\frac{1}{M} \sum_{t=1}^M \mathcal{P}(\mathbf{y} \vert t)\right] \text{d} \mathbf{y} \right\}.  \label{RSpcForm2}
\end{equation}

By incorporating the expression of $\mathcal{P}(\mathbf{y} \vert n)$, we have:
\begin{equation}
  \int \mathcal{P}(\mathbf{y} \vert n) \log_2 \mathcal{P}(\mathbf{y} \vert n) \text{d} \mathbf{y} = -N_\text{R}\log_2(\pi e) - \log_2(|\bm\Sigma_n|).
  \label{RSpcForm2_Part1}
\end{equation}

Moreover, since $\log_2(\cdot)$ is a concave function, the following inequality can be yielded via a direction application of Jensen's inequality:
\begin{equation}
\arraycolsep=1.0pt\def\arraystretch{1.3}
  \begin{array}{rcl}
  && \displaystyle \int \mathcal{P}(\mathbf{y}\vert n) \log_2 \left[\frac{1}{M} \sum_{t=1}^M \mathcal{P}(\mathbf{y} \vert t)\right] \text{d} \mathbf{y}\\
  &\le& \displaystyle \log_2 \left[\frac{1}{M}\sum_{t=1}^M \int \mathcal{P}(\mathbf{y} \vert n) \mathcal{P}(\mathbf{y} \vert t) \text{d} \mathbf{y}\right] = -N_\text{R} \log_2 \pi + \log_2 \left[ \sum_{t=1}^M \frac{|\bm\Sigma_t + \bm\Sigma_n|^{-1}}{M} \right].
  \end{array}
\label{RSpcForm2_Part2}
\end{equation}

By substituting (\ref{RSpcForm2_Part1}) and (\ref{RSpcForm2_Part2}) into (\ref{RSpcForm2}), a lower bound of $I(\mathbf{y}; m)$ is thus given as follows:
\begin{equation}
  I_\text{LB}(\mathbf{y}; m) = \log_2 M - N_\text{R} \log_2 e - \frac{1}{M} \sum_{n=1}^M \log_2 \sum_{t=1}^M \frac{\left|\bm\Sigma_n\right|}{\left|\bm\Sigma_n + \bm\Sigma_t\right|}.
  \label{RSpcLB}
\end{equation}

Finally, by substituting (\ref{RSpcLB}) and (\ref{RSym}) into (\ref{RTotal}), the closed-form lower bound $R_\text{LB}(\mathbf{H}, \mathcal{D}, \mathbf{A})$ in Theorem \ref{theorem1} is thus yielded, which completes the proof.
\end{IEEEproof}

\section{Proof of Proposition \ref{proposition1}}\label{AppB}
\begin{IEEEproof}
According to the derivations in Appendix \ref{AppA}, the closed-form lower bound can be decomposed as:
\begin{equation}
  R_\text{LB}(\mathbf{H}, \mathcal{D}, \mathbf{A}) = I(\mathbf{y}; \mathbf{x}\vert m) + I_\text{LB}(\mathbf{y}; m),
\end{equation}
where $I(\mathbf{y}; \mathbf{x}\vert m)$ and $I_\text{LB}(\mathbf{y}; m)$ are given by (\ref{RSym}) and (\ref{RSpcLB}), respectively. Since the derivation of $I(\mathbf{y}; \mathbf{x} \vert m)$ is accurate, according to (\ref{RSym}), we thus seek to derive the value of $I_\text{LB}(\mathbf{y}; m)$, when an asymptotically high or low SNR value $\rho / \sigma_\text{N}^2$ is invoked.

\textit{Case \uppercase\expandafter{\romannumeral1} (asymptotically high SNR)}: We seek to prove that the following limits hold:
\begin{equation}
  \displaystyle \frac{\left| \bm\Sigma_n \right|}{\left| \bm\Sigma_n + \bm\Sigma_t \right|} \xrightarrow[]{\rho/\sigma_\text{N}^2 \rightarrow \infty}
  \begin{cases}
    0, & n \ne t,\\
    2^{-N_\text{R}}, & n = t.
  \end{cases}
  \label{AppB_Eq0}
\end{equation}

The case of $n = t$ can be readily proved. We now focus on the case of $n \ne t$. On the one hand, when $n \ne t$, the following derivations hold with an asymptotically high SNR:
\begin{equation}
\arraycolsep=1.0pt\def\arraystretch{1.8}
  \begin{array}{rcl}
  \left|\bm\Sigma_n + \bm\Sigma_t\right| &=& \displaystyle \left(2\sigma_\text{N}^2\right)^{N_\text{R}} \cdot \left|\mathbf{I}_{N_\text{R}} + \frac{\rho_\text{S}}{2} \mathbf{HA}\left(\mathbf{C}_n\mathbf{D}_n\mathbf{D}_n^H\mathbf{C}_n^H + \mathbf{C}_t\mathbf{D}_t\mathbf{D}_t^H\mathbf{C}_t^H\right) \mathbf{A}^H \mathbf{H}^H\right| \\
  &=& \displaystyle (2\sigma_\text{N}^2)^{N_\text{R}} \cdot \left|\mathbf{I}_{2N_\text{S}} + \frac{\rho_\text{S}}{2} \mathbf{Q}_{nt}^H \mathbf{Q}_{nt}\right| \\
  &\overset{\text{(a)}}{\approx}& \left(\rho / \sigma_\text{N}^2\right)^{2N_\text{S}} \cdot \sigma_\text{N}^{2N_\text{R}} \cdot 2^{N_\text{R} - 2N_\text{S}} \cdot N_\text{S}^{-2N_\text{S}} \cdot \left| \mathbf{Q}_{nt}^H \mathbf{Q}_{nt}\right|, \\
  \end{array}
  \label{AppB_Eq1}
\end{equation}
where $\rho_\text{S} \triangleq \rho / (\sigma_\text{N}^2 N_\text{S})$, $\mathbf{Q}_{nt} \triangleq [\mathbf{HAC}_n\mathbf{D}_n, \mathbf{HAC}_t\mathbf{D}_t]$, and (a) is obtained by assuming $\rho_\text{S} \gg 1$. Note that, since $N_\text{S} \le \text{rank}(\mathbf{H})$ holds according to (\ref{rankRequirement}), we thus have $\left| \mathbf{Q}_{nt}^H \mathbf{Q}_{nt}\right| > 0$, when $n \ne t$. On the other hand, we have:
\begin{equation}
\arraycolsep=1.0pt\def\arraystretch{1.3}
  \begin{array}{rcl}
  \left|\bm\Sigma_n \right| &=& \sigma_\text{N}^{2N_\text{R}} \left|\mathbf{I}_{N_\text{R}} + \rho_\text{S} \mathbf{HAC}_n\mathbf{D}_n \mathbf{D}_n^H\mathbf{C}_n^H\mathbf{A}^H\mathbf{H}^H\right| \\
  &\overset{\text{(a)}}{\approx}& \left(\rho / \sigma_\text{N}^2\right)^{N_\text{S}} \cdot \sigma_\text{N}^{2N_\text{R}} \cdot N_\text{S}^{-N_\text{S}} \cdot \left|\mathbf{D}_n^H\mathbf{C}_n^H\mathbf{A}^H\mathbf{H}^H \mathbf{HAC}_n\mathbf{D}_n\right|,
  \end{array}
  \label{AppB_Eq2}
\end{equation}
where (a) is again obtained by assuming $\rho_\text{S} \gg 1$. Comparing (\ref{AppB_Eq1}) to (\ref{AppB_Eq2}), it can be observed that, with an asymptotically high SNR, $\left|\bm\Sigma_n + \bm\Sigma_t\right|$ scales linearly with $(\rho / \sigma_\text{N}^2)^{2N_\text{S}} \cdot \sigma_\text{N}^{2N_\text{R}}$, while $\left|\bm\Sigma_n \right|$ only scales linearly with $\left(\rho / \sigma_\text{N}^2\right)^{N_\text{S}} \cdot \sigma_\text{N}^{2N_\text{R}}$. Therefore the limits in (\ref{AppB_Eq0}) can be proved. Based on (\ref{AppB_Eq0}), we thus have:
\begin{equation}
  I_\text{LB}(\mathbf{y}; m) \xrightarrow[]{\rho/\sigma_\text{N}^2 \rightarrow \infty} \log_2 M + N_\text{R} \left(1 - \log_2 e\right).
  \label{AppB_Lim1}
\end{equation}

\textit{Case \uppercase\expandafter{\romannumeral2} (asymptotically low SNR)}: In this case, we have $\bm\Sigma_n \approx \sigma_\text{N}^2 \mathbf{I}_{N_\text{R}}$ for $n \in \{1, 2, \ldots, M\}$. Hence we have:
\begin{equation}
  I_\text{LB}(\mathbf{y}; m) \xrightarrow[]{\rho/\sigma_\text{N}^2 \rightarrow 0} N_\text{R}(1 - \log_2 e).
  \label{AppB_Lim2}
\end{equation}

However, since the random input $m$ is drawn from $m \in \{1, 2, \ldots, M\}$ with equal probability, thus the following limits should hold:
\begin{equation}
  I(\mathbf{y}; m) \rightarrow
  \begin{cases}
    \log_2 M, & \text{if } \rho/\sigma_\text{N}^2 \rightarrow \infty, \\
    0, & \text{if } \rho/\sigma_\text{N}^2 \rightarrow 0.
  \end{cases}
  \label{AppB_Lim3}
\end{equation}

Comparing (\ref{AppB_Lim3}) against (\ref{AppB_Lim1}) and (\ref{AppB_Lim2}), it can thus be seen that a constant shift $N_\text{R}(1 - \log_2 e)$ exits between the asymptotic values of $I(\mathbf{y}; m)$ and $I_\text{LB}(\mathbf{y}; m)$, which completes the proof.
\end{IEEEproof}

\section{Proof of Proposition \ref{proposition2}}\label{AppC}
\begin{IEEEproof}
According to \cite{boyd2004convex}, to prove that $R_\text{LB}$ is concave with respect to $\bm\lambda$, it suffices to check that the following function of one variable $s$, i.e.
\begin{equation}
  R_\text{LB}(s) \triangleq R_\text{LB}(\bm\lambda + s \bm\omega),
\end{equation}
is concave with respect to $s$ for any given $\bm\lambda, \bm\omega \in \mathbb{R}_{MN_\text{S} \times 1}$.

Let $\bm\lambda = [\bm\lambda_1^T, \ldots, \bm\lambda_M^T]^T$, $\bm\omega = [\bm\omega_1^T, \ldots, \bm\omega_M^T]^T$ with $\bm\lambda_m, \bm\omega_m \in \mathbb{R}_{N_\text{S} \times 1}$ denoting the $m$-th sub-vectors of $\bm\lambda$ and $\bm\omega$, we can thus define the function $f_n: \mathbb{R} \rightarrow \mathbb{R}$ as follows:
\begin{equation}
  f_n(s) \triangleq \log_2 \sum_{t=1}^M |\bm\Sigma_n + \bm\Sigma_t|^{-1},
\end{equation}
with $\bm\Sigma_n$ given by ($n = 1, 2, \ldots, M$):
\begin{equation}
  \bm\Sigma_n = \sigma_\text{N}^2 \mathbf{I}_{N_\text{R}} + \frac{\rho}{N_\text{S}} \mathbf{HAC}_n \left( \bm\Lambda_n + s \bm\Omega_n \right)\mathbf{C}_n^H \mathbf{A}^H \mathbf{H}^H,
\end{equation}
where $\bm\Lambda_n = \text{diag}(\bm\lambda_n)$ and $\bm\Omega_n = \text{diag}(\bm\omega_n)$. Therefore $|\bm\Sigma_n + \bm\Sigma_t|^{-1}$ can be re-formulated as:
\begin{equation}
\arraycolsep=1.0pt\def\arraystretch{1.3}
  \begin{array}{rcl}
  \left|\bm\Sigma_n + \bm\Sigma_t\right|^{-1} &=& \displaystyle  \left|2\sigma_\text{N}^2 \mathbf{I}_{N_\text{R}} + \frac{\rho}{N_\text{S}} \mathbf{HA} \left( \mathbf{C}_n \bm\Lambda_n\mathbf{C}_n^H + \mathbf{C}_t \bm\Lambda_t \mathbf{C}_t^H \right) \mathbf{A}^H \mathbf{H}^H + \right. \\
  && \displaystyle \left.s \cdot \frac{\rho}{N_\text{S}} \mathbf{HA} \left( \mathbf{C}_n \bm\Omega_n\mathbf{C}_n^H + \mathbf{C}_t \bm\Omega_t \mathbf{C}_t^H \right) \mathbf{A}^H \mathbf{H}^H \right|^{-1}.
  \end{array}
  \label{AppC_Eq0}
\end{equation}

According to (\ref{AppC_Eq0}), $|\bm\Sigma_n + \bm\Sigma_t|^{-1}$ is therefore log-convex with respect to $s$. As summation preserves the log-convexity \cite{boyd2004convex}, $\sum_{t=1}^M |\bm\Sigma_n + \bm\Sigma_t|^{-1}$ is thus also log-convex over $s$, which proves the convexity of $f_n(s)$ with respect to $s$. Finally, since
\begin{equation}
  R_\text{LB}(s) = \log_2 \frac{M}{\left(e \sigma_\text{N}^2\right)^{N_\text{R}}} - \frac{1}{M} \sum_{n=1}^M f_n(s),
\end{equation}
$R_\text{LB}(s)$ is thus verified to be concave with respect to $s$, which completes the proof.
\end{IEEEproof}

\end{document}